\documentclass[prb,letter,twocolumn,
showpacs,lengthcheck,superscriptaddress]{revtex4}
\usepackage{graphicx}
\usepackage{amsmath,amssymb}
\usepackage{bm}% bold math
\usepackage{ifthen}
\usepackage{dcolumn}% Align table columns on decimal point

\def\mc{\multicolumn}

\begin{document}
\title{ Multiple charging of InAs/GaAs quantum dots by electrons or holes:
addition energies and ground-state configurations}
\author{Lixin He}
%\author{Gabriel Bester}
\author{Alex Zunger}
\affiliation{National Renewable Energy Laboratory, Golden CO 80401}
\date{\today}

\begin{abstract}

Atomistic pseudopotential plus configuration interaction calculations 
of the energy needed to charge dots by either electrons or holes 
are described, and contrasted with the widely used, but highly simplified
two-dimensional
parabolic effective mass approximation (2D-EMA). Substantial discrepancies are
found, especially for holes, regarding the stable electronic configuration and
filling sequence which defies both Hund's rule and the Aufbau principle. 

\end{abstract}

\maketitle

\section{Introduction}

One of the most spectacular aspects of quantum-dot physics is that dots can be
controllably charged by either electrons or holes and that one can measure,
for each of the many charged-states, both the electronic spectrum
and the charging energies. This is afforded  either by injecting from a
tip of a scanning tunneling microscope, \cite{banin99} or
by various gate structures. \cite{kastner93,tarucha96,drexler94,fricke96,
miller97, kouwenhoven97,bodefeld99, regelman01,
warburton00, reuter05} 
Since the energy scale of both single-particle levels 
and Coulomb interactions in quantum dots (QDs) 
($10^{-4}$ - $10^{-2}$ Ry) 
are a few order of magnitudes smaller
than those of the real atoms ( $\sim$ 1 Ry),
dots can be loaded by as many as six \cite{drexler94} to 
ten \cite{banin98} electrons in colloidal \cite{banin98}
and self-assembled \cite{drexler94,bodefeld99,regelman01} dots having
confining dimension of $\leq$ 50 \AA, 
and up to hundreds of electrons in larger 500 \AA\,
electrostatically confined dots. \cite{tarucha96,kouwenhoven97,kastner93}
The ``charging energy'' $\mu(N)$ is the energy needed to add a
carrier to the dot that is already loaded by $N-1$ carriers, 
\begin{equation}
\mu(N)=E(N)-E(N-1) \, ,
\label{eq:charging}
\end{equation}
where $E(N)$ is the correlated, many-body total energy of the ground state
of the $N$-particle dot. The ``addition energy'' $\Delta(N-1,N)$ 
(analogous to the difference
between ionization potential and electron affinity)
indicates how much more energy is needed 
to add the $N$th carrier compared to the energy needed to add the 
$(N-1)$th carrier:
\begin{eqnarray}
\Delta(N-1,N) &=& \mu(N)-\mu(N-1) \nonumber \\
              &=& E(N)-2E(N-1)+E(N-2) \, .
\label{eq:addition}
\end{eqnarray}
The typical electron addition energies for electrostatic dots,
\cite{tarucha96,kouwenhoven97,kastner93} are about 1 - 8 meV, 
and the stable spin-configuration
follows the rules of atomic physics; that is, the 
{\it s, p, d,} ... shells are occupied in successive order with no holes 
left behind
(Aufbau principle) and with maximum spin\cite{tarucha96} (Hund's rule). 
Recently, it became possible to load and measure electrons
\cite{drexler94,bodefeld99} and holes
\cite{bodefeld99,reuter04,reuter05} 
into much smaller,
% ($\simeq$ 200 $\times$ 40 \AA),
epitaxially grown self-assembled dots of InGaAs/GaAs, 
where electron addition energies are about
10 - 60 meV, \cite{drexler94,bodefeld99} 
and hole addition energies are between 10 - 30 meV.\cite{reuter04,reuter05} 
More interestingly, while electrons still follow the Aufbau principle, 
recent hole charging experiment \cite{reuter05} 
show that holes have unusual charging patterns   
that defy the Aufbau principle and Hund's rule.

Despite the importance of the dot charging problem and the great success 
achieved in experimentally recording the charging spectra, 
the theoretical understanding of charging and addition
energies is still preliminary. 
Most theoretical works in the area were based on particle-in-a-potential model,
\cite{jacak_book,wojs96,warburton98,rontani98}
neglecting interband (e.g., $\Gamma$-$\Gamma$) coupling, 
intervally ($\Gamma$-$X$-$L$) coupling effect, 
and the true atomistic symmetry
(e.g., $C_{2v}$ for lens of zincblende material) 
which is lower than the shape symmetry.
The most often used potential in such approaches is the
2-dimensional (2D) parabolic form, in which all of
the above noted electronic structure effects are replaced by
an effective mass approximation (EMA).
In this 2D-EMA model,\cite{jacak_book,warburton98} 
the single particle levels have equal spacing, which equals 
a harmonic oscillator frequence $\omega$. Because of the simplicity
of the model, all the Coulomb integrals can be
related  analytically \cite{warburton98} to a single $s$ orbital Coulomb energy
$J_{ss}$, and therefore the 
addition energies are determined entirely
by  $\omega$ and $J_{ss}$. 
Although the 2D-EMA model can be attractive because of its 
algebraic simplicity,
and availability of fitting parameters,
actual self-assembled dots are significantly different 
from the description of 
EMA model, manifesting inter-band coupling and inter-valley coupling,
strain effects, low atomistic symmetry, 
as well as specific band offset profiles,
all neglected by the 2D-EMA. 
It indeed has been recently shown \cite{he05d} that hole ground state
configurations predicted 
by the 2D-EMA model are qualitatively different from 
those measured by hole charging experiments.\cite{reuter04,reuter05}

An atomistic pseudopotential description of electronic structure
effects can be used instead of 2D-EMA to calculate charging energies.
\cite{williamson00, bester03b,bester05a,franceschetti00,he05d}
Here, we show that such an atomistic theory correctly
reproduces the many-particle configurations as well as addition 
spectra for carriers in self-assembled quantum dots.
We study systematically 
the electronic structure of
self-assembled InGaAs/GaAs quantum dots, provide detailed information
on the electron and hole single-particle spectrum, many-particle 
charging and addition spectrum, as well as ground state configurations 
using single-particle pseudo-potential 
and many-particle configuration interaction (CI) methods.

The rest of the paper is arranged as follows. 
In Sec.~\ref{sec:background}, we introduce the basic concepts of charging and 
addition energies, and show how to calculate these quantities in the 
single-particle pseudopotential plus many-particle CI scheme.
In Sec.~\ref{sec:results}, we give detailed results calculated from
pseudopotential-CI scheme the
single-particle levels, Coulomb integrals and
the ground state configurations as well as
the addition energies. We contrast these results 
with 2D-EMA model. We summarize in Sec.~\ref{sec:summary}.

\section{Theory of dot charging and addition energies}
\label{sec:background}

\subsection{General equation for dot charging
in the configuration-interaction approach}

The calculation of the total-energy $E(N)$ of $N$-particle dot requires
obtaining first the single-particle states from an effective Schr\"{o}dinger
equation, and then the many-particle state from a many-particle treatment. The
first step is formulated as,
\begin{equation}
\left[ -{1 \over 2} \nabla^2 
+ V_{\rm ext}({\bf r}) + V_{\rm scr}({\bf r}) \right] \psi_i({\bf r})
=\epsilon_i \;\psi_i({\bf r}) \; ,
\label{eq:schrodinger}
\end{equation}
where $V_{\rm ext}({\bf r})$ is the external (``bare'') potential experienced
by the electrons or holes, and $V_{\rm scr}({\bf r})$ is the screening
response. The single-particle orbital $\{\psi_i\}$ and energies
$\{\epsilon_i\}$ are used in the second step to construct the many-particle
wave functions $\{\Psi\}$ and energies $\{E\}$ from, 
\begin{equation}
E(N)=\langle \Psi_N |H |\Psi_N\rangle \;
\label{eq:tot-eng}
\end{equation}
where, the many-body Hamiltonian is,
\begin{eqnarray}
H=&&\sum_{i\sigma} \epsilon_{\alpha}\psi^{\dag}_{i\sigma} 
\psi_{i\sigma} \nonumber \\
&&+ {1\over2} \sum_{ijkl} \sum_{\sigma_1,\sigma_2} 
\sum_{\sigma_3,\sigma_4}
\Gamma^{i\sigma_1,j\sigma_2}_{k\sigma_3,l\sigma_4} 
\psi^{\dag}_{i\sigma_1} \psi^{\dag}_{j\sigma_2}
\psi_{k\sigma_3}\psi_{l\sigma_4}\, ,
\label{eq:ham}
\end{eqnarray}
and,
\begin{widetext}
\begin{equation}
\Gamma^{i\sigma_1,j\sigma_2}_{k\sigma_3,l\sigma_4} 
=\sum_{{\bf s}_1,{\bf s}_2}\int\int d{\bf r}d{\bf r'}\; 
{\psi^*_{i\sigma_1}({\bf r},{\bf s}_1) \psi^*_{j\sigma_2} ({\bf r'},{\bf s}_2) 
\psi_{k\sigma_3}({\bf r'},{\bf s}_2) \psi_{l\sigma_4} ({\bf r},{\bf s}_1) 
\over \epsilon({\bf r}- \bf{r'}) |{\bf r} -{\bf r'}|} \, ,
\label{eq:int_h}
\end{equation} 
\end{widetext}
are the screened Coulomb and exchange integrals.
In the above Eqs.~(\ref{eq:ham}) and (\ref{eq:int_h}), 
``$\sigma$'' is a pseudospin
index, i.e., an index of Kramers degenerate states, 
while ``${\bf s}$'' is the {\it intrinsic} electronic spin. 
For {\it electrons} in InAs/GaAs QDs, the spin-orbit interactions 
is extremely small and can be neglected. In this case,
the pseudospin $\sigma$ and intrinsic electronic spin ${\bf s}$ are equivalent.
However, for {\it holes}, which have a mixture of 
heavy-, (H) light-hole (LH) and
split-off character, an eigenstate of $\sigma$ has both
${\bf s}=\uparrow$ and ${\bf s}=\downarrow$ components.  
The $N$-particle wave functions can be solved using e.g. configuration 
interaction (CI) method, \cite{szabo_book} 
by expanding the $N$-electron wave function
in a set of Slater determinants,
$|\Phi_{e_1,e_2,\cdots,e_N}\rangle
=\psi^{\dag}_{e_1}\psi^{\dag}_{e_2}\cdots\psi^{\dag}_{e_N}|\Phi_0\rangle$,
where $\psi^{\dag}_{e_i}$
creates an electron in the state $e_i$ .
The $\nu$-th many-particle wave function is then the linear combination of
the determinants, 
\begin{equation}
|\Psi^{(\nu)}_N\rangle=\sum_{e_1,e_2,\cdots,e_N}
A_{\nu}(e_1,e_2,\cdots,e_N)\;|\Phi_{e_1,e_2,\cdots,e_N}\rangle \; .
\label{eq:coeff}
\end{equation}
Once $|\Psi_N\rangle $ is known, we can then calculate the 
corresponding total energies for the ground states as well as excited
states using Eq. (\ref{eq:tot-eng}). 
Once we solve the CI problem, we get the order of total CI energy for
various holes or
electron configurations, so we  can see if Hund's rule or the Aufbau principle
or spin-blockade occurs. For example, {\it Hund's rule} states
that degenerate single-particle
levels are occupied with maximum number of unpaired electrons,
while the Aufbau principle states, 
non-degenerate single-particle levels are occupied
in order of increasing single-particle energy.  

We construct all possible Slater determinants corresponding to 
$N$ electrons or $N$ holes [i.e., we ignore the excitonic 
(electron+hole) excitations], using only the bound states 
of the dots, (i.e., we neglect all continuum states).
The underlying electrons that are not
considered explicitly by this approach are
represented by the dielectric screening function 
$\epsilon({\bf r}- \bf{r'})$ in Eq. (\ref{eq:int_h}).

\subsection{The Hartree-Fock equations for charging and addition energies} 

\begin{figure}
\includegraphics[width=2.5in,angle=0]{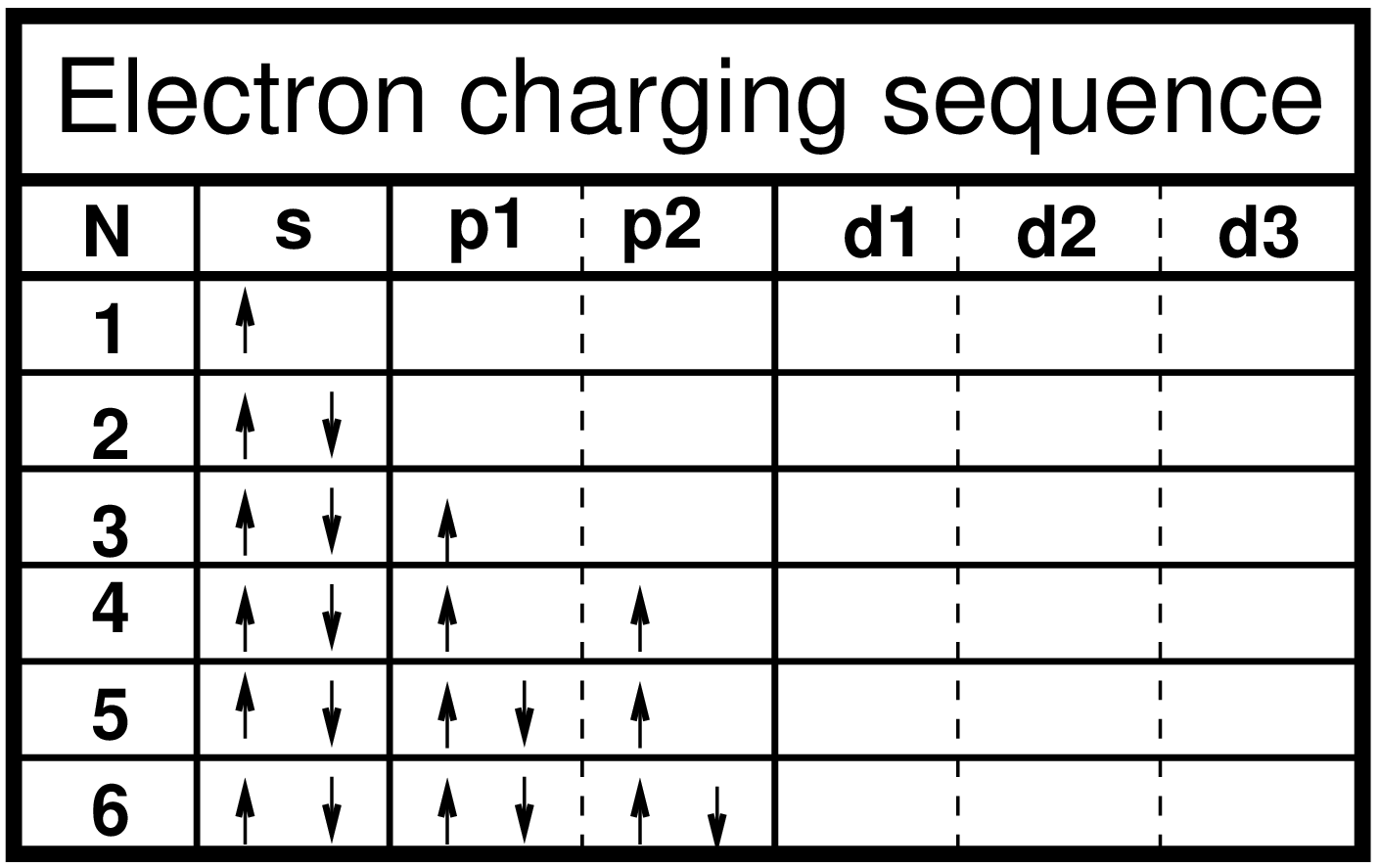}
\caption{ The sequence of electron configurations for filling 
$N$ electrons to the quantum dots according to
both 2D-EMA model and pseudopotential calculations.
Note that the $p$ levels are split (into $p_1$ and $p_2$),
so are the $d$ levels ($d_1$, $d_2$, $d_3$). 
}
\label{fig:config_e}
\end{figure}

The addition energies at CI level can be written as the
Hartree-Fock (HF) addition energies plus the correlations, i.e.,
\begin{equation}
\Delta_{\rm CI}(N-1,N)=\Delta_{\rm HF}(N-1,N)
+\Delta_{\rm corr.}(N-1,N) \, ,
\label{eq:correct-CI}
\end{equation}  
where $\Delta_{\rm corr.}$ is the correlation energy correction
to the addition energy calculated in HF. Since the HF equations are
used by many experimentalists
to deduce Coulomb energies,\cite{bodefeld99,reuter05,tarucha96}
we review it here.
In the Hartree-Fock approximation, where the effect of 
correlations is neglected but the direct Coulomb and exchange interactions
are retained, simple expressions can be derived for the addition energies.
The total energy of $N$ electrons is simply,
\begin{equation}
E_{\rm HF} = \sum_{i\sigma}^{\rm occ.} \epsilon_{i\sigma} 
+ \sum_{i\sigma,j\sigma'}^{\rm occ.} 
(J_{i\sigma,j\sigma'} - K_{i\sigma,j\sigma'})\;,
%\delta_{\sigma\sigma'})\;,
\label{eq:HF-eng}
\end{equation}
where, $i$ is the single-particle level index of all occupied states,
and $\epsilon_i$ is the corresponding single-particle energy.
The $J_{i\sigma,j\sigma'}=\Gamma^{i\sigma,j\sigma'}_{j\sigma',i\sigma}$ and
$K_{i\sigma,j\sigma'}=\Gamma^{i\sigma, j\sigma'}_{i\sigma,j\sigma'}$
in Eq. (\ref{eq:HF-eng}) are Coulomb and exchange integrals, respectively. 
Since the spin index ``$\sigma$'' is not
an actual electronic spin, but
rather an index for two Kramers degenerate states, in principle
the exchange integrals $K_{i\sigma,j\sigma'}$ are not simply 
diagonal in $\sigma$, $\sigma'$,  as has been widely used in the 
dot charging literature.
\cite{warburton98,tarucha96,wojs96,reuter05}
%However, we can diagonal $K_{i\sigma,j\sigma'}$
%in the $\sigma$, $\sigma'$ basis and choose the largest eigenvalue
%as $K_{i,j}$.
% 
However, adopting the literature approximation, 
$K_{i\sigma,j\sigma'}$=$K_{i,j}\,\delta_{\sigma\sigma'}$
and considering  the $s$ and two $p$ orbitals, 
$\epsilon_{p_1} <\epsilon_{p_2}$
and assuming the particle filling order follows Hund's rule, as shown in 
Fig.\ref{fig:config_e},
the total energies for $N$=1, 2, 3, 4 electrons in the Hartree-Fock
approximation are
\begin{widetext}
\begin{eqnarray}
E_{\rm HF}(1)&=&\epsilon_s \; , \nonumber \\
E_{\rm HF}(2)&=&2\epsilon_s + J_{s,s}  \; , \nonumber \\
E_{\rm HF}(3)&=&2\epsilon_s + \epsilon_{p_1} 
+ J_{s,s} +2J_{s,p_1}-K_{s,p_1}  \; , \nonumber \\
E_{\rm HF}(4)&=&2\epsilon_s + \epsilon_{p_1} +\epsilon_{p_2}
+ J_{s,s} +2J_{s,p_1}+2J_{s,p_2}+J_{p_1,p_2} 
-K_{s,p_1}-K_{s,p_2} - K_{p_1,p_2} \; \nonumber \\
E_{\rm HF}(5)&=&2\epsilon_s + 2\epsilon_{p_1} +\epsilon_{p_2}
+J_{s,s}+4J_{s,p_1}+2J_{s,p_2}+J_{p_1,p_1}+2J_{p_1,p_2}
- 2K_{s,p_1}-K_{s,p_2} - K_{p_1,p_2} \;\nonumber \\
E_{\rm HF}(6)&=&2\epsilon_s + 2\epsilon_{p_1} +2\epsilon_{p_2}
+J_{s,s}+4J_{s,p_1}+4J_{s,p_2}+J_{p_1,p_1}+J_{p_2,p_2}+ 4J_{p_1,p_2}
\;\nonumber \\
& &- 2K_{s,p_1}-2K_{s,p_2} -2K_{p_1,p_2} \;.
\end{eqnarray}
We can then readily calculate the charging energies 
Eq. (\ref{eq:charging}) in this approximation,
\begin{eqnarray}
\mu_{\rm HF}(1)&=&\epsilon_s \; , \nonumber \\
\mu_{\rm HF}(2)&=&\epsilon_s + J_{s,s}  \; , \nonumber \\
\mu_{\rm HF}(3)&=&\epsilon_{p_1}+ 2J_{s,p_1}-K_{s,p_1}  \; , \nonumber \\
\mu_{\rm HF}(4)&=&\epsilon_{p_2}+2J_{s,p_2}+J_{p_1,p_2} -K_{s,p_2}
-K_{p_1,p_2}  \; , \nonumber \\
\mu_{\rm HF}(5)&=&\epsilon_{p_1}+2J_{s,p_1}+J_{p_1,p_1}
+J_{p_1,p_2}-K_{s,p_1}\; , \nonumber \\
\mu_{\rm HF}(6)&=&\epsilon_{p_2}+2J_{s,p_2}+2J_{p_1,p_2}+J_{p_2,p_2}-K_{s,p_2}
-K_{p_1,p_1} \; .
\end{eqnarray}
Similarly, we can calculate the addition energies 
Eq. (\ref{eq:addition}) as follows,
\begin{eqnarray}
\Delta_{\rm HF}(1,2)&=&J_{s,s} \; , \nonumber \\
\Delta_{\rm HF}(2,3)&=&(\epsilon_{p_1}-\epsilon_s) 
+2J_{s,p_1}- J_{s,s} -K_{s,p_1}  \; , \nonumber \\
\Delta_{\rm HF}(3,4)&=& (\epsilon_{p_2}-\epsilon_{p_1}) 
+ 2J_{s,p_2}-2J_{s,p_1} +J_{p_1,p_2} -K_{s,p_2} 
+K_{s,p_1} -K_{p_1,p_2}  \;, \nonumber \\ 
\Delta_{\rm HF}(4,5)&=& (\epsilon_{p_1}-\epsilon_{p_2}) +2J_{s,p_1}
-2J_{s,p_2}+J_{p_1,p_1}-K_{s,p_1}+K_{s,p_2}+K_{p_1,p_2}\;, \nonumber \\ 
\Delta_{\rm HF}(5,6)&=& (\epsilon_{p_2}-\epsilon_{p_1}) +2J_{s,p_2}
-2J_{s,p_1}+J_{p_1,p_2}-J_{p_1,p_1}+J_{p_2,p_2} \nonumber \\ 
&& +K_{s,p_1}-K_{s,p_2}-K_{p_1,p_2}\;.
\label{eq:delta_HF}
\end{eqnarray}
\end{widetext}
From above equations, we see that to predict $\Delta(N-1,N)$ for
the first two cases ($N$=3), we need to know {\it 4 parameters} ($J_{ss}$,
$J_{sp_1}$, $K_{sp_1}$, $\epsilon_{p_1}-\epsilon_s$), 
but for $N$=6, we need to know {\it 9 parameters}.
Because of the large number of parameters needed, the analysis of
charging effects in the literature resort to additional 
approximations aimed at reducing the
number of parameters using simplified effective mass models.
Equations (\ref{eq:delta_HF}) are going to be used below to contrast the
charging spectra deduced from simplified literature
models {\it vs.} our more complete treatment.

\subsection{Atomistic treatment of the single-particle problem}
\label{sec:background:single}
The most general treatment of the single-particle problem of 
Eq.(\ref{eq:schrodinger}) describes
both $V_{\rm ext}$ and $V_{\rm scr}$ atomistically, much in the same way as
molecules are treated quantum-mechanically. In this description 
$V_{\rm ext} ({\bf r})$ is a superposition of the ionic potential of
individual atoms of type $\alpha$ at lattice site $n$,
\begin{equation}
V_{\rm ext}({\bf r}) =\sum_{\alpha}\sum_n v_{\rm ion}^{(\alpha)} 
({\bf r}-\tau_{\alpha}-{\bf R}_n)\; ,
\end{equation}
and $v_{\rm ion}^{(\alpha)}$ is $-Z_{\alpha}/r$ in an all-electron 
(core+valence) treatment
(where $Z_\alpha$ is the atomic number) or $v_{\rm ps}^{(\alpha)}({\bf r})$ in
the pseudopotential (valence-only) scheme 
(where $v_{\rm ps}^{(\alpha)}({\bf r})$ is the
ionic pseudopotential. The potential $V_{\rm ext}$ naturally contains the
correct point-group symmetry of the object, through the atomic position
vectors $\{\tau_{\alpha}, {\bf R}_n\}$, and includes atomic relaxation if
appropriate (again, through the atomic positions), as well as chemical
inhomogeneity (alloying) or surface-passivation effect. The screening response
$V_{\rm scr}({\bf r})$ of Eq.~(\ref{eq:schrodinger}) 
is in general a functional of the density-matrix 
$\rho({\bf r}, {\bf r}')$ and can be described e.g. via Hartree-Fock or the
density functional theory, both requiring a self-consistent (iterative)
solution to Eq.~(\ref{eq:schrodinger}). These approaches are currently limited
to small dots, relative to the 10$^3$ - 10$^5$ atom dots which charging
experiments exist. Furthermore, LDA suffer 
from the famous ``LDA error'', whereby the band gap and
effective masses are badly underestimated.  
Higher-order method such as GW approximation or time-depend DFT have yet to
demonstrate applicability to large dots for which high-quality experiments
exit.  

An approximation to the screening 
$V_{\rm scr}[\rho({\bf r}, {\bf r}')]$, which allows calculation on large dots,
and fixes the LDA error, 
is provided by the ``screened
pseudopotential approach'', where it is assumed that $V_{\rm scr}({\bf r})$ can
be described as a superposition of screening potentials $v_{\rm
  scr}^{(\alpha)}({\bf r})$ of the individual atoms, and lumping together 
$v_{\rm ion}^{(\alpha)} ({\bf r}) + v_{\rm scr}^{(\alpha)} ({\bf r})$
to yield a screened atomic pseudopotential $v_{\rm epm}^{(\alpha)} ({\bf r})$,
such that,
\begin{equation}
V_{\rm ext} ({\bf r}) + V_{\rm scr} ({\bf r}) = V_{\rm so}({\bf r})+
\sum_{\alpha}\sum_{n} v_{\rm epm}^{(\alpha)} 
({\bf r}-\tau_{\alpha}-{\bf R}_n)   \; .
\label{eq:epm}
\end{equation}
Here, $v_{\rm epm}^{(\alpha)} ({\bf r})$ is determined semi-empirically.
Unlike the classic empirical pseudopotential method,~\cite{cohen66} which
fitted only to eigenvalues, here we require
that when Eq.~(\ref{eq:schrodinger}) 
is applied to the underlying {\it bulk} periodic 
solids containing atom $\{\alpha\}$, $\{v_{\rm epm}^{(\alpha)}\}$ reproduce
the {\it measured} band energies, effective-mass tensors, deformation potentials,
and the single-particle wave functions have a large overlap with the
corresponding LDA wave functions.\cite{wang95,fu97b} In Eq. (\ref{eq:epm}), 
a nonlocal potential $V_{\rm so}({\bf r})$ is also 
added to the total potential 
to represent the spin-orbit interaction.
In our approach, the potential of an As atom depends on the number of Ga and In atoms around
it as
\begin{equation}
v_{\rm As} ({\rm Ga}_n{\rm In}_{4-n}) = {n\over 4} v_{\rm As}({\rm Ga}) 
+ {4-n\over 4} v_{\rm As}({\rm In})\, ,
\label{eq:As-pp}
\end{equation}
where $n$ is the number of Ga atoms around the As atom.
In this atomistic approach, one assume that $v_{\rm epm}^{(\alpha)} ({\bf r})$
is transferable to different environments. Note that a fixed $v_{\alpha}$
is a good approximation if the
dot has no {\it free} surfaces (as is the case in self-assembled dots, where
only a strained interface between chemically-similar materials is present). For surface
atoms in free-standing dots, a separate 
$v_{\rm epm}^{(\alpha)} ({\bf r})$ is fitted \cite{fu97c} 
to LDA {\it surface} calculations. For InAs/GaAs dots, we use the pseudopotentials
of Ref.\onlinecite{williamson00}. These pseudopotentials have been tested not only
for the InAs and GaAs binaries, but also for alloys and superlattices of the
corresponding ternaries.~\cite{williamson00}

Once $\{ v_{\rm epm}^{(\alpha)} ({\bf r}) \}$ is known, one can solve 
Eq.~(\ref{eq:schrodinger})
for the bulk solid, quantum wells superlattices, quantum-wires or quantum dots
by adopting a supercell approach where the respective objects are placed.
In our case, Eq.~(\ref{eq:schrodinger}) is solved using the ``linear 
combination of Bloch
bands'' (LCBB) method,\cite{wang99b}
where the wave functions $\psi_i$ are expanded as, 
\begin{equation}
\psi_{i}({\bf r}) =\sum_{n,{\bf k}}\sum_{\lambda}
C_{n,{\bf k}}^{(\lambda)}\;
\phi_{n,{\bf k},\tensor{\epsilon}}^{(\lambda)}({\bf r})\; .
\label{eq:lcbb}
\end{equation}
In the above equation,
$\{\phi_{n,k,\tensor{\epsilon}}^{(\lambda)}({\bf r})\}$ are the bulk
Bloch orbitals of band index $n$
and wave vector ${\bf k}$ of material $\lambda$ (= InAs, GaAs), strained
uniformly to strain $\tensor{\epsilon}$.
The inclusion of stain-dependent basis functions improves their variational
flexibility.
We use  $\tensor{\epsilon}=0$ for the
(unstrained) GaAs matrix material, and an average $\tensor{\epsilon}$ value 
from VFF for the strained dot material (InAs).
For the InAs/GaAs system, we use $n=8$ (including spin) 
for electron states on a
6$\times$6$\times$16 k-mesh. Note that the potential 
$V_{\rm ext} ({\bf r}) +V_{\rm src} ({\bf r})$ contain full strain effects
through the use of {\it relaxed} atomic positions, in addition to 
the explicit strain~\cite{williamson00} and 
alloy composition~\cite{magri02} dependence.

In the atomistic approach to the single-particle problem, one 
includes (i) multi-band coupling [different $n$ in Eq. (4)]; (ii) inter-valley
coupling [different bulk $k$-points in Eq. (4)]; (iii) spin-orbit coupling
[$V_{\rm so}$ in Eq. (\ref{eq:epm})]; 
(iv) the proper strain profile [by relaxing $\{\tau_{\alpha}, {\bf
 R}_n\}$ in Eq. (\ref{eq:epm}) 
to minimize strain]; (v) realistic chemical profile
[distributing the species $\{ \alpha \}$ as in a random alloy, 
\cite{williamson00} or interdiffused interfaces \cite{magri02}]. 
The ensuing single-particle orbitals 
$\{\psi_i\}$ transform like the representations of the point-group created by
the ionic positions. These underlying atomistic structures could
break the symmetry 
represented by the macroscopic shape of the quantum dots. 
For example, a lens-shaped dot has a macroscopic cylindrical symmetry with
[110] and [$\bar{1}$10] being equivalent, 
but if the dot is made of a zincblende material, the real symmetry is $C_{2v}$,
where, [110] and [$\bar{1}$10] are not equivalent.
Yet the continuum models do not
``see'' the atomistic symmetry (the ``farsightedness effect''\cite{zunger02}). 
Therefore, as discussed in Ref.~\onlinecite{he04a}, in reality 
the atomistic wave function need not to be simple ``pure'' $s$-like 
or $p$-like.
As a result, the Coulomb energy  $J_{i,j}$  and exchange energy $K_{i,j}$
obtained with atomistic
single-particle orbital $\{\psi_i\}$ do not have simple 
relationships \cite{warburton98} as predicted by continuum 2D-EMA model.
The deviations of the atomistic calculated $J$s from the simplified 2D-EMA
ones are going to lead to new physical behavior
(e.g., new ground state symmetry of the many-particle state), 
as illustrated below.

\subsection{Continuum treatment of the single-particle problem: EMA}

It is sometimes customary \cite{warburton98,reuter05,tarucha96,dekel00}
to avoid an atomistic description of the
single-particle problem in favor of a single-band particle-in-a-box model. In
this approach, one sets $V_{\rm scr}$=0 and replaces $V_{\rm ext}$ of Eq. (3)
by a pure external potential, describing the {\it macroscopic shape} of the
object, e.g., a box represents
a quantum well, or a sphere with finite or infinite barriers represents a 
quantum dot. 
\cite{jacak_book,bimberg_book,warburton98} 
In some cases, this $V_{\rm ext}$ is calculated ``realistically''
from a combination of band offset, the gate potential and the
ionized impurities,\cite{fonseca98,bednarek03}
but it is treated nevertheless as a macroscopic {\it external} field. 
Under this approximation, simple results can be obtained for 
cylindrical QDs, where the angular momentum  $L$ and $L_z$ are good quantum 
numbers. 
The electron (hole) single-particle levels have well defined shell structures,
with non-degenerate $s$ shell,
two-fold degenerate $p$ shell, and three-fold degenerate $d$ shell, etc. 
Furthermore, for parabolic confining potentials,
the single-particle levels have equal spacing between two adjacent shells,
e.g. $\epsilon_{p}-\epsilon_{s}$=$\epsilon_{d}-\epsilon_{p}$.
Another reason for the attractiveness of the 
single-band particle-in-a-box approach to the single-particle problem
[Eq.(\ref{eq:schrodinger})], is that the many-particle problem 
[Eq.(\ref{eq:ham})] 
becomes simple. For example, the quantum Monte Carlo 
(QMC) approach
is currently applied {\it only} within the
single-band particle-in-a-box approach 
for such large objects as self-assembled quantum dots. 
\cite{Shumway01} 
In a 2D-EMA model all the Coulomb integrals 
needed for charging and addition energy can be all related 
to \cite{warburton98} $J_{ss}$, 
\begin{eqnarray}
J_{\alpha\beta}&=&c^{(J)}_{\alpha\beta}\, J_{ss} \nonumber \\
K_{\alpha\beta}&=&c^{(K)}_{\alpha\beta}\, J_{ss}\, .
\label{eq:J-Jss}
\end{eqnarray}
For example, $J_{sp}=3/4 J_{ss}$,
$K_{sp}=1/4 J_{ss}$.
Therefore the charging/addition energies and ground 
state configurations are totally determined by $J_{ss}$ and
the single-particle energy spacing $\omega$.

However, real self-assembled quantum dots grown {\it via} the 
Stranski-Krastanov techniques, are not well-described 
by the single-band particle-in-a-box approaches, despite the great popularity
of such approaches in the experimental literatures.
\cite{warburton98,reuter05,dekel00,wojs96} 
%In particular, the EMA neglect inter-band coupling, inter-valley coupling, the
%explicit effects of strain and chemical inhomogeneity. 
%This leads to 
The model contains significant 
quantitative errors \cite{wang99c} and also 
qualitative errors, whereby 
cylindrically symmetric dots are deemed to have, by symmetry, no
fine-structure splitting, 
no polarization anisotropy, 
and no splitting of $p$ levels and $d$ levels, all being a
manifestation of the ``farsightedness effect''.\cite{zunger02} 

\section{results}
\label{sec:results}

\begin{table*}
\caption{Summary of the pseudopotential-calculated
single-particle level spacing (in meV)
of In$_{1-x}$Ga$_x$As/GaAs quantum dots
of different heights, base sizes, and
Ga compositions. $e_i$ and $h_i$ are the $i$-th electron and
hole single-particle energy levels. $E_X$ is the lowest exciton
energy.}
\label{tab:single}
\vskip 0.2cm
\begin{tabular}
%{lddddddddd}
{p{2.4cm} p{1.5cm} p{1.5cm} p{1.5cm} p{1.5cm} p{1.5cm} 
p{1.5cm}p{1.5cm}p{1.5cm}p{1.5cm} }
%{dddddddddd}
\hline\hline
Height (nm) &2.5 & 3.5 & 2.5& 3.5 & 3.5 & 3.5 & 3.5 
& 2.5 & 3.5\\
Base (nm)   & 20  & 20 & 25  &25 &25 &25 &25 &27.5 &27.5 \\
Ga comp.  &  0     &   0    &  0     &0 & 0.15  & 0.3    & 0.5&  0& 0 \\
\hline
$e_{p_1}-e_s$ & 77.0 & 72.0 & 60.8 & 57.0& 51.5& 44.2 & 33.2 & 54.9 & 51.0\\
$e_{d_1}-e_{p_2}$ & 79.7 & 73.6 & 64.0  & 58.7 & 50.7 & 43.2 & 18.8 & 57.9 
& 52.3\\
$e_{p_2}-e_{p_1}$ & 2.5  & 3.7  & 1.6 & 2.1 &2.5 & 1.2 & 0.8 & 1.1  & 2.0  \\
$e_{d_2}-e_{d_1}$ & 3.1  & 2.8  & 1.2 & 1.4 & 4.0 & 1.3 &4.1 & 0.8  & 1.0 \\
\hline
$h_s-h_{p_1}$& 17.8 & 10.4 & 17.4  & 11.3 & 13.1& 13.7 & 12.8 &17.0 & 11.3  \\
$h_{p_1}-h_{p_2}$& 10.9 & 11.3 & 7.1 & 9.5& 7.1 & 5.0 & 3.3 & 5.8  & 7.9 \\
$h_{p_2}-h_{d_1}$ & 4.5 & 3.4  & 8.3 & 2.4 &6.4 & 8.6 & 9.4 & 9.4  & 3.9 \\
\hline
%$e_0-h_0$ & \\
$E_X$ & 1080 &1035 &1042 & 996 & 1095 & 1188 &1297 &1028 & 981 \\
\hline\hline
\end{tabular}
\end{table*} 

Using the pseudopotential approach for single-particle and configuration
interaction approach for the many-particle step, we 
studied the electron or hole addition energy spectrum 
up to 6 carriers in lens-shaped InAs dots embedded in a
GaAs matrix. We study dots of three different base size,
$b$= 20 nm, 25 nm and 27.5 nm, and for each base size, 
two heights, $h$=2.5 nm and 3.5 nm. 
To study the alloy effects, we also calculated
the addition spectrum for alloy dots In$_{1-x}$Ga$_x$As/GaAs of
$h$/$b$=3.5/25 nm dots, with Ga composition $x$=0, 0.15, 0.3 and 0.5.
In this section, we give detailed results 
of the single particle energy levels and Coulomb integrals, 
and the addition energy as well as ground state configurations.
We also compare the results with what can be 
expected from 2D-EMA model.

\subsection{Single-particle level spacing: Atomistic {\it vs.} 2D-EMA
  description }

\begin{figure}
\includegraphics[width=2.6in,angle=0]{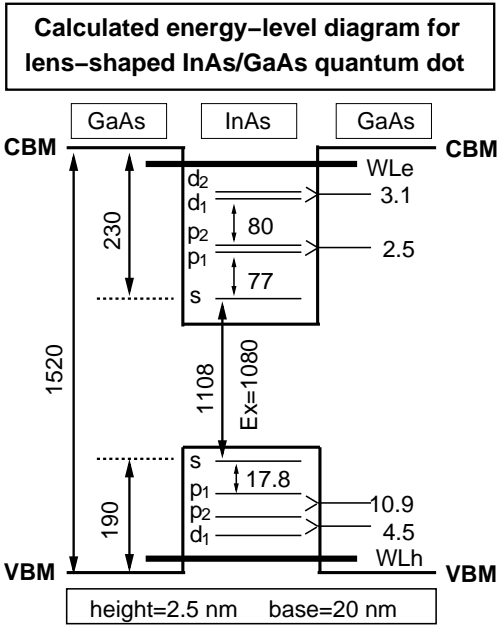}
\caption{The schematic energy-level diagram (in meV) 
of a pure lens-shaped InAs/GaAs quantum dots, 
with height $h$=2.5 nm and base $b$=20 nm.
WLe and WLh denote the wetting layer
energy levels for electrons and holes respectively.
The CBM and VBM correspond to the  
conduction band minima and valence band maxima of
(unstrained) bulk GaAs. $E_x$ is the excitonic transition energy.
} 
\label{fig:eng-level}
\end{figure}

\subsubsection{Electron levels} 

We depict in Fig.~\ref{fig:eng-level} the calculated energy-level 
diagram of a pure lens-shaped InAs/GaAs quantum dot, 
with height $h$=2.5 nm and base $b$=20 nm. 
Figure~\ref{fig:eng-level} shows that the electron confinement energy is 230
mev, somewhat larger than the hole confinement energy (190 meV). The $p$ levels
are split as are the $d$ levels, even though the dot has macroscopic
cylindrical symmetry (see below).  

The pseudopotential calculated electron single-particle energy spacings 
are summarized in Table~\ref{tab:single} for QDs
of different heights, bases, and alloy compositions. 
Table~\ref{tab:single} gives
the fundamental exciton energy $E_X$ calculated from CI approach
for each dot.
These exciton energies are between 980 meV - 1080 meV for pure 
InAs/GaAs dots, and can be as large as 1297 meV 
for In$_{1-x}$Ga$_x$As/GaAs alloy dots.
This range
agrees very well with experimental results for these classes of dots, ranging 
from 990 meV to 1300 meV.
\cite{schmidt96,schmidt98,bodefeld99,reuter05}

(a) {\it $s$-$p$ and $p$-$d$ energy spacing}:
From Table~\ref{tab:single}, we see that for electrons in the lens-shaped
dot, the $s$-$p$ energy level spacing $\delta_{sp}=\epsilon_p-\epsilon_s$ 
and $p$-$d$ energy level spacing $\delta_{pd}=\epsilon_d-\epsilon_p$  
are nearly equal, as assumed by the 2D harmonic
model. The energy spacing $\delta_{sp}$ 
and $\delta_{pd}$ range from 50 - 80 meV 
(Fig.~\ref{fig:eng-level}), 
depending on the dot geometries.
%For electrons, the energy spacings are sensitive 
%to the base size of the dots,
The electron energy spacings decrease with increasing QD
base sizes.
The electron energy spacing of alloy dots are much smaller 
than those in pure InAs/GaAs
QDs, because of reduced confinement.
For Ga rich dots ($x$=0.3 - 0.5), the single-particle energy level spacings 
range from 30 - 45 meV. These values agree with the
infrared absorption measurements \cite{drexler94,fricke96}
of intra-band transitions of 
alloy InGaAs QDs, which give $\delta_{sp} \sim$ 41 - 45 meV.
When the Ga composition reaches $x$=0.5, 
the $s$-$p$ energy level spacing $\delta_{sp}$ 
becomes significantly different
from the $p$-$d$ energy level spacing
$\delta_{pd}$, thus deviating from
harmonic potential approximation.

\begin{figure}
\includegraphics[width=2.8in,angle=0]{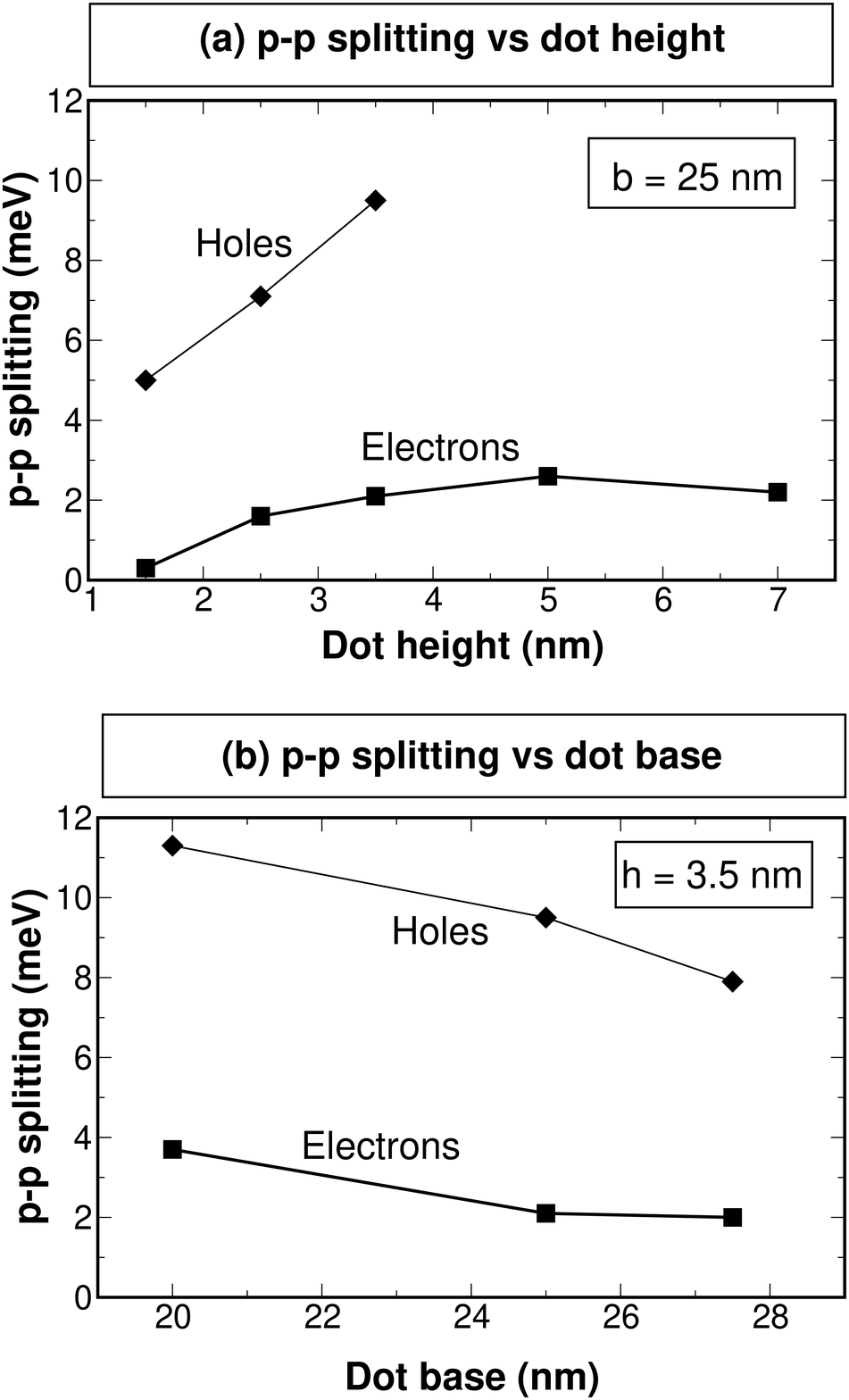}
\caption{Pseudopotential-calculated
$p$ level splitting of electrons
and holes 
vs. (a) dot height
and (b) dot base for InAs/GaAs dots. 
}
\label{fig:dpp}
\end{figure}

(b) {\it Shell definition}: 
Figure~\ref{fig:eng-level} shows that the
energy levels of electrons in a lens-shaped dot
have well defined $s$, $p$ $d$ shell structure. 
However, while effective mass and ${ k\cdot p}$ models predict degenerate
$p$ and $d$ levels, for cylindrically-symmetric (e.g. lens-shaped) QDs,  
atomistic calculations show that even in perfect lens-shaped dots,
the $p$-$p$ and $d$-$d$ levels are split by
2 - 4 meV (Fig.~\ref{fig:eng-level}) 
due to the actual $C_{2v}$ symmetry.
We denote the two $p$ levels as $p_1$ and $p_2$, and similarly,
the three $d$ levels as $d_1$, $d_2$ and $d_3$, 
in increasing order of energy.  
The results listed in Table~\ref{tab:single} show 
that $\delta_{pp}=\epsilon_{p_2}-\epsilon_{p_1}$ 
and $\delta_{dd}=\epsilon_{d_2}-\epsilon_{d_1}$  
are very sensitive to the
aspect ratio of the dots while not being
very sensitive to the alloy compositions.  
Figure \ref{fig:dpp} depicts 
$\delta_{pp}=\epsilon_{p_2}- \epsilon_{p_1}$ 
vs. dot heights [Fig.\ref{fig:dpp}(a)] 
and bases sizes [Fig.\ref{fig:dpp}(b)]. 
In general, we see that $\delta_{pp}$
increases with increasing dot height, 
and it decreases with increasing dot base size. 

\subsubsection{Hole levels}

In contrast to electrons, hole single-particle levels (Table \ref{tab:single})
display a much more complicated behavior that is  
totally beyond the EMA description.

(a) {\it $s$-$p$ spacing}: 
As one can see from Table \ref{tab:single},
the hole $s$-$p$ energy spacing ranges from 10 - 18 meV
for dots of sizes we studied. These
energy spacings are considerably smaller than those
of electrons.  
The first confined hole state is found to 
be about 190 meV above the VBM of bulk GaAs, for the 
pure, $h/b$=2.5/20 nm dot
(Fig.~\ref{fig:eng-level}).   
Unlike the case for 
electrons, the energy spacing between hole 
$s$ and $p$ levels
depends strongly on the height of the dots, \cite{footnote0}
while being
relatively insensitive to the base size of the dots.

(b) {\it Shell definition}: 
The well-defined $s$, $p$, $d$ shell-structure for electrons
does not exist for holes (Fig.~\ref{fig:eng-level}),
as the $p_1$-$p_2$ and $d_1$-$d_2$-$d_3$ 
splitting are much larger than those for
electrons.\cite{footnote1} 
The hole $p$-level splittings are also shown 
in Fig.~\ref{fig:dpp} for different
dot heights [Fig.\ref{fig:dpp}(a) ] 
and bases sizes [Fig.\ref{fig:dpp}(b)].
For the smallest dots, 
$h$/$b$ = 2.5/25 nm, the $p$ splitting is about 11 meV
(Table \ref{tab:single}), 
more than 3 times the value for electrons. This splitting
is about half of the hole $s$-$p$ energy spacing. 
Note that the pseudopotential calculated $p$-$p$ 
splitting is much lager
than 1.3 meV given by the ${k \cdot p}$ method 
(which includes piezoelectric effect).~\cite{sheng05}
For taller dots, the $p$-$p$ splittings are even larger. 
As a consequence, the 
$p_2$ levels are energetically very close to the $d_1$ levels, 
leading, as we will see below,
to a nontrivial charging pattern that breaks Hund's rule and the Afubau
principle. \cite{he05d} 

(c) {\it Wave function characters}:
An analysis of the wave function show that these levels
have somewhat mixed $S$, $P$ or $D$ characters. For example, for a
lens-shaped InAs/GaAs QD
of height=3.5 nm and base=25 nm, the ``$s$ level'' has 92\% $S$ character,
and the two ``$p$ levels'', $p_1$ and $p_2$, have
 86\% $P$ character respectively (see 
Table II of Ref.~\onlinecite{he04a}). 
Therefore, we can still 
label the single-particle levels as $s$, $p_1$, $p_2$, etc. 
These single-particle levels, $s$, $p_1$ and
$p_2$, do not have pure HH character
either, being instead 91\%, 86\%, and 92\% HH-like respectively.
As the aspect ratio height/base increases, 
the mixture of angular momentum and  HH and LH characters becomes stronger.
For example, as shown in Ref.~\onlinecite{narvaez05},
for a InGaAs/GaAs alloy dot, with 25.2 nm in base and 2 nm in
height, $s$ level has 90\% $S$ character,
while both $p_1$ and $p_2$ levels have 84\% $P$ character.
When the height of the dot increases to 7.5 nm (with fixed base
size), the leading angular momentum characters for these three
levels are 84\% $S$,  78\% $P$ and 75 \% $P$ respectively.
Similarly, for the flat dot (2 nm in height), 
the mixture of LH state character is about 4 - 9 \%,
but increases to 11 - 17 \% for the tall dot (7.5 nm in height).

\begin{table*}
\caption{{\it Electron-electron}
Coulomb energies $J_{\alpha\beta}/J_{ss}$ and 
exchange energies $K_{\alpha\beta}/J_{ss}$ 
calculated numerically 
from atomistic pseudopotential theory for pure InAs/GaAs quantum dots
compared with results obtained by the 2D-EMA model. The dots have
base $b$= 25 nm with different heights.}
\label{tab:JK-e}
\vskip 0.2cm
\begin{tabular}{p{2.4cm} p{2.4cm} p{1.5cm} p{1.5cm} p{1.5cm} p{1.5cm} p{1.5cm} }
%{ccddddd}
\hline\hline
          & \mc{1}{l}{ 2D-EMA} & \mc{5}{c}{Atomistic} \\
Height (nm)  &   & 1.5  & 2.5 & 3.5 & 5.0  & 7.0 \\
\hline
$J_{ss}$ (meV)        &  -     & 25.1 & 24.3  &  22.6 & 21.3  & 19.7\\
$J_{sp}$/$J_{ss}$     & 0.75  & 0.80   &  0.83   & 0.84  & 0.85   & 0.86\\ 
$J_{sd_1}$/$J_{ss}$   & 0.59  & 0.67   &  0.73   & 0.75  & 0.76   & 0.77\\
$J_{sd_2}$/$J_{ss}$   & 0.69  & 0.63   &  0.72   & 0.75  & 0.75   & 0.77\\
$J_{p_1p_1}$/$J_{ss}$ & 0.69  & 0.78   &  0.84   & 0.84  & 0.86   & 0.87\\
$J_{p_1d_1}$/$J_{ss}$ & 0.60  & 0.61   &  0.73   & 0.75  & 0.77   & 0.78\\
$J_{p_1d_2}$/$J_{ss}$ & 0.58  & 0.61   &  0.71   & 0.73  & 0.74   & 0.75\\
$J_{d_1d_1}$/$J_{ss}$ & 0.57  & 0.55   &  0.71   & 0.75  & 0.76   & 0.78\\
$J_{d_2d_2}$/$J_{ss}$ & 0.60  & 0.59   &  0.70   & 0.73  & 0.75   & 0.76\\
$J_{d_1d_2}$/$J_{ss}$ & 0.53  & 0.53   &  0.64   & 0.67  & 0.68   & 0.70\\
\hline
$K_{sp_1}$/$J_{ss}$    & 0.25  &  0.20  & 0.22     & 0.23 & 0.22  & 0.21\\
$K_{sd_1}$/$J_{ss}$    & 0.09  &  0.12  & 0.09     & 0.09 & 0.09  & 0.09\\ 
$K_{sd_2}$/$J_{ss}$    & 0.19  &  0.06  & 0.08     & 0.09 & 0.08  & 0.08\\
$K_{p_1p_2}$/$J_{ss}$  & 0.19  &  0.06  & 0.07     & 0.07 & 0.07  & 0.07\\ 
$K_{p_1d_1}$/$J_{ss}$  & 0.24  &  0.09  & 0.19     & 0.20 & 0.21  & 0.20\\ 
$K_{p_1d_2}$/$J_{ss}$  & 0.11  &  0.13  & 0.16     & 0.17 & 0.17  & 0.17\\ 
\hline\hline
\end{tabular}
\end{table*}

\begin{table*}
\caption{{\it Hole-hole}
Coulomb energies $J_{\alpha\beta}/J_{ss}$ and 
exchange energies $K_{\alpha\beta}/J_{ss}$ calculated numerically 
from atomistic pseudopotential theory for pure InAs/GaAs quantum dots
compared with results obtained by 2D-EMA model. The dots have
base $b$= 25 nm with different heights.}
\label{tab:JK-h}
\vskip 0.2cm
\begin{tabular}{p{2.4cm} p{2.4cm} p{1.5cm} p{1.5cm} p{1.5cm} p{1.5cm} p{1.5cm}}
%{lcddddd}
\hline\hline
          & \mc{1}{l}{ 2D-EMA} & \mc{5}{c}{Atomistic} \\
Height (nm)  &   &  1.5  & 2.5  & 3.5
& 5.0  & 7.0 \\
\hline
$J_{ss}$  (meV)   &   -     &  27.2 &  25.1 &  20.4 & 14.3   & 14.3 \\
$J_{sp}/J_{ss}$      & 0.75  & 0.79  &  0.80   &  0.85  & 0.94   &  1.00\\ 
$J_{sd_1}/J_{ss}$   & 0.59  & 0.70  &  0.70   &  0.70  & 0.87   &  0.73 \\
$J_{sd_2}/J_{ss}$   & 0.69  & 0.80  &  0.81   &  0.76  & 0.95   &  1.02\\
$J_{p_1p_1}/J_{ss}$  & 0.69  & 0.73  &  0.74   &  0.79  & 0.92   &  0.73\\
$J_{p_1d_1}/J_{ss}$  & 0.60  & 0.67  &  0.68   &  0.70  & 0.87   &  0.70\\
$J_{p_1d_2}/J_{ss}$  & 0.58  & 0.68  &  0.70   &  0.74  & 0.92   &  0.70\\
$J_{d_1d_1}/J_{ss}$  & 0.57  & 0.65  &  0.65   &  0.67  & 0.87   &  0.80\\
$J_{d_2d_2}/J_{ss}$  & 0.60  & 0.71  &  0.72   &  0.72  & 0.94   &  0.80\\
$J_{d_1d_2}/J_{ss}$  & 0.53  & 0.63  &  0.64   &  0.69  & 0.86    & 0.77\\
\hline
$K_{sp_1}/J_{ss}$    & 0.25  & 0.22  &  0.24   & 0.27   & 0.38  & 0.79 \\
$K_{sd_1}/J_{ss}$    & 0.09  & 0.10  &  0.10   & 0.06   & 0.13  & 0.14\\ 
$K_{sd_2}/J_{ss}$   & 0.19  & 0.14  &  0.20   & 0.12   & 0.31  & 0.13\\
$K_{p_1p_2}/J_{ss}$  & 0.19  & 0.14  &  0.14   & 0.16   & 0.36  & 0.12\\ 
$K_{p_1d_1}/J_{ss}$  & 0.24  & 0.23  &  0.24   & 0.26   & 0.21  & 0.14\\ 
$K_{p_1d_2}/J_{ss}$ & 0.11  & 0.10  &  0.12   & 0.14   & 0.20  & 0.12\\ 
\hline\hline
\end{tabular}
\end{table*}

\begin{figure}
\includegraphics[width=3.in,angle=0]{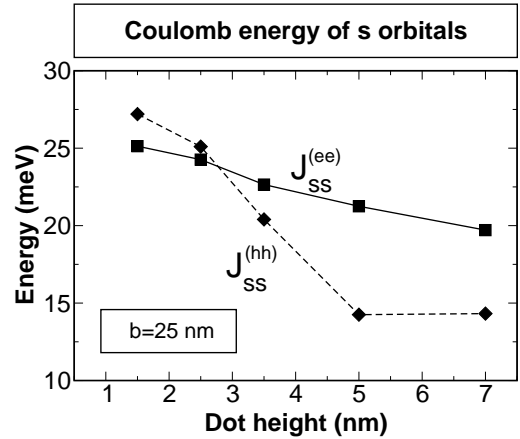}
\caption{ Comparison of the electron-electron (ee) 
and hole-hole (hh) Coulomb energies of $s$ orbitals 
$J_{ss}$ for lens-shaped dots with different heights. 
}
\label{fig:Jss}
\end{figure}

\subsection{Coulomb integrals: atomistic {\it vs} 2D-EMA description}
\label{sec:coulomb}

Another piece of information that decides the addition energies 
[Eq.~({\ref{eq:addition})]
is the Coulomb integral between the particles [Eq.~({\ref{eq:delta_HF})].
We list the pseudopotential-calculated
Coulomb energies of $s$ orbitals $J_{ss}$ for electrons in the first
row of Table~\ref{tab:JK-e} and for holes in the first row
of Table~\ref{tab:JK-h}.
For electrons, $J_{ss}^{(ee)} \sim $ 22 - 25 meV, and for holes 
$J_{ss}^{(hh)} \sim $ 20 - 27 meV for typical dots. These numbers can be directly
compared with experimental value from electron/hole charging experiments, 
since $J_{ss} \approx \Delta (1,2)$ [Eq. (\ref{eq:delta_HF})]. The typical
experimental values of $J_{ss}$ for electrons is about 
 \cite{drexler94,warburton98,bock02} 19 - 27 meV
and for holes \cite{warburton98,bock03,reuter05} 20 - 25 meV.  
We plot in Fig.\ref{fig:Jss} $J_{ss}$ for electrons and holes vs. 
height for base $b$ = 25 nm dots. For flat dots,
the electron-electron Coulomb energy $J_{ss}^{(ee)}$ is smaller than that of
holes $J_{ss}^{(hh)}$. 
However, $J_{ss}^{(ee)}$ is larger than $J_{ss}^{(hh)}$ for taller dots. 
The crossover is at about 2.5 nm
for the $b$ = 25 nm dots. 
Note that
in our calculation, the two nearly degenerate electron $p$ orbitals 
$p_1$ and $p_2$, are spatially almost orthogonal to each other. 
However, in the simple 2D-EMA model,\cite{warburton98} 
the two degenerate $p$
orbitals $p_+=p_x+ i\,p_y$ and $p_-= p_x-i\,p_y$
have same spatial function differing only by a phase factor.
As a result, the  exchange interaction between $p_1$ and $p_2$  
is much smaller than that of $p_+$ and $p_-$.
Furthermore, the simple 2D-EMA model predicts
$J_{p_+p_-}=J_{p_+p_+}$, which is not true in the atomistic description, 
where $J_{p_1p_2}$ is much smaller than $J_{p_1p_1}$ and  $J_{p_2p_2}$. 

In a 2D-EMA model, there is only one free parameter in calculating the
Coulomb integrals for each type of
carriers: the effective length $l_e$ for electrons 
and $l_h$ for holes.
This leads to simple geometric relationships as 
illustrated in Eq.~(\ref{eq:J-Jss}). 
These relations are listed in the second column of
Tables~\ref{tab:JK-e} and \ref{tab:JK-h}. 
However, as noted in Sec.~\ref{sec:background:single}, 
a detailed analysis of atomistic wave functions \cite{he04a,narvaez05} 
show that they are not of pure conduction band character for electrons 
or HH, LH characters for holes; 
nor do they have pure $s$, $p$ angular momentum 
characters as predicted by 2D-EMA model. Since 
inter-valley, inter-band coupling and the underlying atomistic symmetries
are also ignored in the 2D-EMA model, 
%which considers only the envelope 
%wave functions.
we might expect that the simple
relations between the Coulomb integrals of Eq.~(\ref{eq:J-Jss})
must be somehow different in the atomistic approach relative to 
those predicted by the EMA, {\it even} for rather flat lens-shaped
dots having parabolic-like energy level pattern, 
$\epsilon_p-\epsilon_s \approx \epsilon_d-\epsilon_p$, as chosen here.

The ratios $J_{\alpha\beta}/J_{ss}$ and $K_{\alpha\beta}/J_{ss}$
derived from atomistic calculations are compared
with the 2D-EMA model relationships 
in Table~\ref{tab:JK-e} for electrons and
in Table~\ref{tab:JK-h} for holes for dots of different
aspect ratios. To visualize the differences, 
we also plot the relative error for Coulomb energies
$(J_{\alpha\beta}/J_{ss})_{\rm atm} -(J_{\alpha\beta}/J_{ss})_{\rm EMA}$ 
and for exchange energies
$(K_{\alpha\beta}/J_{ss})_{\rm atm} -(K_{\alpha\beta}/J_{ss})_{\rm EMA}$ in 
Fig.~\ref{fig:Jss-compare} (a) for
electrons and in  Fig.~\ref{fig:Jss-compare}(b) for holes, 
where ``atm'' means atomistic
results and ``EMA'' means the 2D-EMA model results.  
For flat lens-shaped dots which have a parabolic-like level
spectrum, we see qualitative agreement between the atomistic calculations and
the 2D-EMA model. For such flat dots,
the errors are generally within 20\% of $J_{ss}$.
For taller dots, the agreement is worse, as can be seen from  
Fig.~\ref{fig:Jss-compare}.
For even taller QDs (e.g., height/base $>$ 5/25 nm), 
the large biaxial strain will
develop a ``hole trap'' at the InAs/GaAs interface, which lead to
the hole localization on the interface of the dots.\cite{he04a,narvaez05} 
In these cases, where the hole (envelope) wave functions are totally 
different from those predicted by 2D-EMA model, the relationship between
Coulomb and exchange energies are, accordingly, very different from those of  
2D-EMA model.

\begin{figure}
\includegraphics[width=3.0in,angle=0]{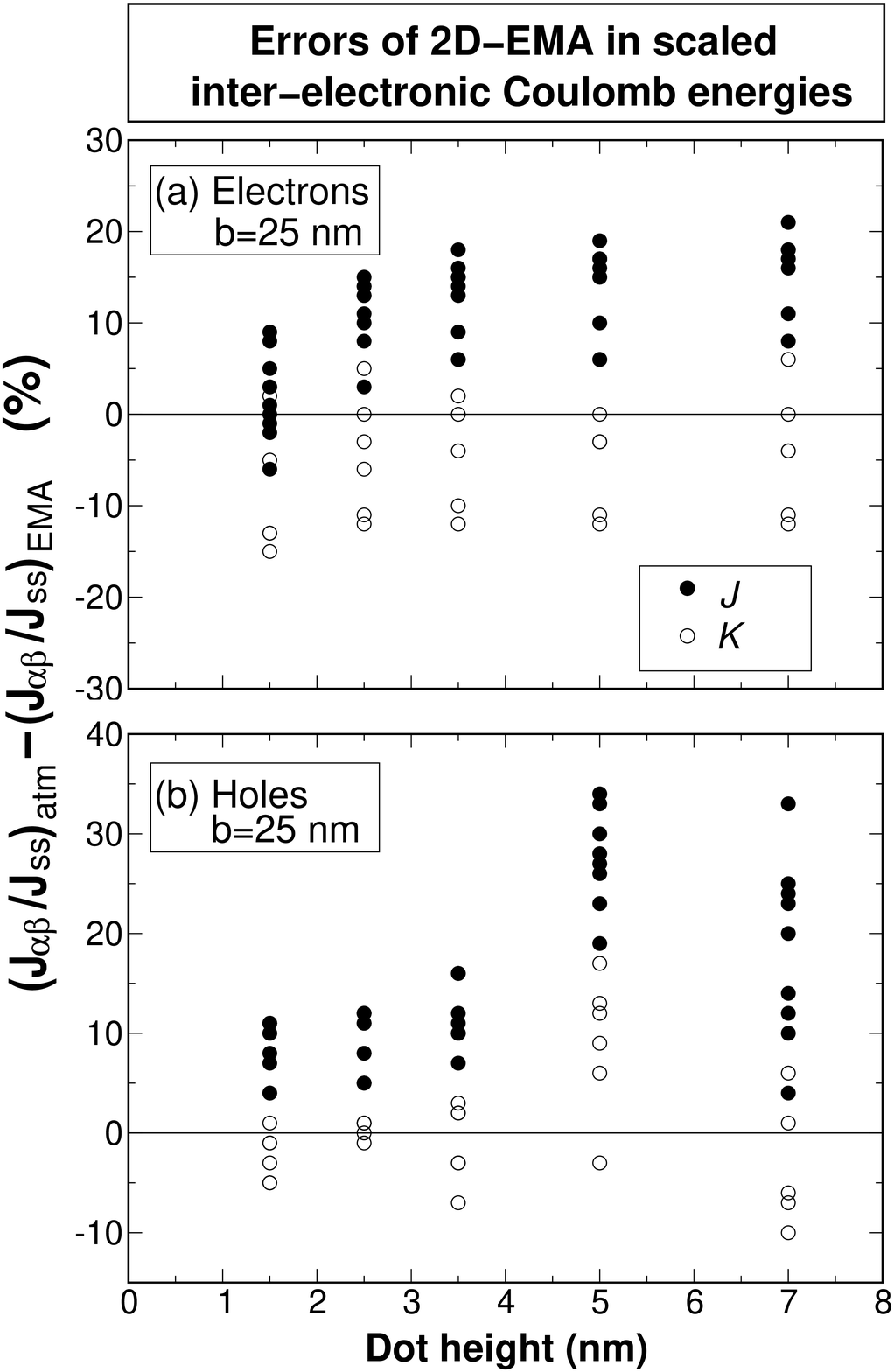}
\caption{The relative error $(J_{\alpha\beta}/J_{ss})_{\rm atm} 
-(J_{\alpha\beta}/J_{ss})_{\rm EMA}$ (closed circles)
and $(K_{\alpha\beta}/J_{ss})_{\rm atm} -(K_{\alpha\beta}/J_{ss})_{\rm EMA}$ 
(open circles) for (a) electrons and (b) holes, {\it vs.} dot height for 
dots of base $b$ = 25 nm. Here, 
``atm'' means atomistic pseudopotential calculation,
and ``EMA'' means the 2D-EMA model.
 }
\label{fig:Jss-compare}
\end{figure}

At first sight, it would seem that
if the values of $J_{ss}$ in the 2D-EMA model are taken to be those
calculated atomistically, 
the Coulomb integrals predicted by 2D-EMA model are 
only a few meV off from those calculated by atomistic theories.  
However, these small differences will change significantly
the electronic phase-diagrams 
of carriers, as we will see below.

\subsection{Ground state configurations: atomistic 
{\it vs.} 2D-EMA description}
\label{sec:configuration}

\begin{figure*}
\includegraphics[width=4.5in,angle=0]{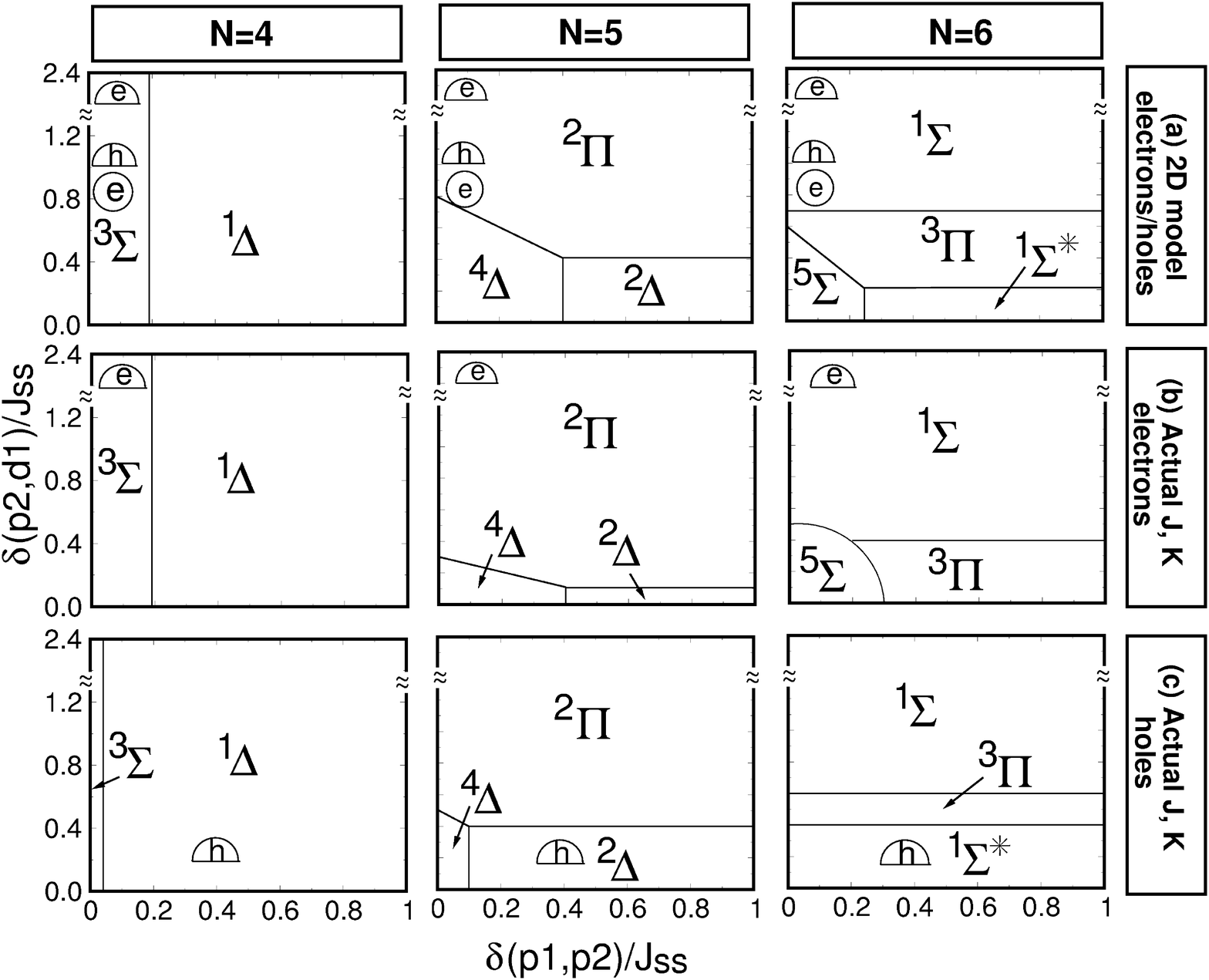}
\caption{The most stable configurations in self-assembled
InAs/GaAs dots, for $N=$ 4, 5, 6 electrons/holes, as a function of the
normalized $p_1$-$p_2$ splitting 
$\delta(p_1,p_2)=\epsilon_{p_2}- \epsilon_{p_1}$ and 
$p_2$-$d_1$ splitting 
$\delta(p_2,d_1)=\epsilon_{d_1}- \epsilon_{p_2}$.
The notation of the configurations are given in 
Eqs. (\ref{eq:phase-4}) - (\ref{eq:phase-6}). 
(a) Using the 2D-EMA model (the 
results apples to both electrons and holes). 
(b) For electrons, using atomistic Coulomb 
and exchange integrals. 
(c) For holes, using atomistic Coulomb and exchange integrals.
The $\delta(p_1,p_2)/J_{ss}$ and $\delta(p_2,d_1)/J_{ss}$
values of actual self-assembled dots are denoted by ``e'' and ``h'' in lenses
for electrons and holes respectively. 
We also show the most stable electron 
configurations in the phase diagrams of a typical
electrostatic dot represented by the circles
for 2D-EMA model. 
}
\label{fig:phase}
\end{figure*}

\subsubsection{Generic phase diagrams for 2D-EMA model}

To calculate the addition energies, we first need to know the spin and
orbital configurations of the few-particle ground state, 
which minimizes the 
energy $E(N)$ of the $N$-particle system.
For an illustration, we first use the ``single configuration approximation'', 
i.e., retain only the Slater determinants
that have the same orbital degree of freedom,
and keep the exchange (spin-spin) coupling
between these Slater determinants.
% (i.e., single-configuration
%approximation)
%to calculate the total energies of all possible
%spin and orbital configurations. 
The ground states are those
configurations that have the lowest total energy.
For electrons, the calculated ground state for $N$=1 - 6
are shown in Fig.~\ref{fig:config_e},
which shows that the electron charging pattern follows Hund's rule.
The ground state spin and orbital configurations for holes are listed 
in Table~\ref{tab:spin-h} 
for $N$= 1 - 6 holes.
In contrast to electrons, the hole charging patterns show complicated
behaviors that defy the Hund's rule and the Aufbau's principle
for most of the cases of $N$=5, 6 holes.\cite{he05d}

To understand the differences between the ground-state configurations of 
electrons and holes and  the 
driving forces for these differences, 
we developed a general phase-diagram approach~\cite{he05d} that 
classifies the many-particle configurations for
electrons and holes in quantum dots in terms of simple electronic and
geometric parameters.
To do so, we calculate for each particle number $N$,
the configuration which minimizes the total-energy under
the single-configuration approximation
at different $p_1$-$p_2$ splitting 
($\delta_{p_1,p_2}=\epsilon_{p_2} -\epsilon_{p_1}$) 
and $p_2$-$d_1$ energy spacing 
($\delta_{p_2,d_1}=\epsilon_{d_1} -\epsilon_{p_2}$)
using Coulomb integrals $J$ and $K$ in units of $J_{ss}$. 
This approach gives a phase diagram as a function of
the parameters $\{ {N; \delta_{p_1,p_2}, \delta_{p_2,d_1} } \}$ 
in units of $J_{ss}$,
which yields for $N$=4 two electronic phases:
\begin{eqnarray}
^3\Sigma &= &(s^\uparrow s^\downarrow)
(p_{1}^{\uparrow})(p_{2}^{\uparrow})\, , \nonumber \\
^1\Delta &=&(s^\uparrow s^\downarrow)
(p_{1}^{\uparrow} p_{1}^{\downarrow})\, .
\label{eq:phase-4}
\end{eqnarray}
Here, we have adopted a spectroscopic
notation for a system with cylindrical symmetry.
For $N=5$, there are three possible phases:
\begin{eqnarray}
^4\Delta &=&(s^\uparrow s^\downarrow)(p_{1}^{\uparrow})
(p_{2}^{\uparrow})(d_1^{\uparrow})\, , \nonumber \\ 
^2\Pi&=&(s^\uparrow s^\downarrow)(p_{1}^{\uparrow}p_{1}^{\downarrow})
(p_2^{\uparrow})\, , \nonumber \\ 
^2\Delta&=&(s^\uparrow s^\downarrow)(p_{1}^{\uparrow}p_{1}^{\downarrow})
(d_{1}^{\uparrow})\, .
\label{eq:phase-5}
\end{eqnarray}
For $N=6$, we find four phases, 
\begin{eqnarray}
^5\Sigma&=&(s^\uparrow s^\downarrow)(p_{1}^{\uparrow})
(p_{2}^{\uparrow})(d_1^{\uparrow})(d_2^{\uparrow}) \, , \nonumber \\ 
^3\Pi&=&(s^\uparrow s^\downarrow)(p_{1}^{\uparrow}p_{1}^{\downarrow})
(p_2^{\uparrow})(d_1^{\uparrow}) \, , \nonumber \\
^1\Sigma &=&(s^\uparrow s^\downarrow)(p_{1}^{\uparrow}p_{1}^{\downarrow})
(p_{2}^{\uparrow}p_{2}^{\downarrow}) \, , \nonumber \\
^1\Sigma^* &=&(s^\uparrow s^\downarrow)(p_{1}^{\uparrow}p_{1}^{\downarrow})
(d_{1}^{\uparrow}d_{1}^{\downarrow})\, .
\label{eq:phase-6}
\end{eqnarray}
We first apply this approach to a 2D-EMA model.
We relax the restrictions in 2D-EMA model of degenerate shells 
($\epsilon_{p_1}$=$\epsilon_{p_2}$, 
$\epsilon_{d_1}=\epsilon_{d_2}=\epsilon_{d_3}$ ) 
and of equidistant
shells ($\epsilon_p-\epsilon_s=\epsilon_d-\epsilon_p$) 
and allow 
$\delta_{p_1,p_2}$ and $\delta_{p_2,d_1}$ to vary.
The resulting phase diagrams are shown in Fig. \ref{fig:phase}(a) 
for $N$=4, 5 , 6.

\subsubsection{Ground states for specific dots in the 2D-EMA model}

To decide which of these phases is a ground state for a given 
dot, we need to know in Fig. \ref{fig:phase}(a)
the actual value of $\delta_{p1,p2}/J_{ss}$ and $\delta_{p2,d1}/J_{ss}$
for this realistic dot. 
For electrons in self-assembled dots, the single-particle energy spacing
is usually more than twice the Coulomb energy,\cite{drexler94,warburton98}
i.e., $\delta_{p2,d1} > 2 J_{ss}$. 
For holes, $\delta_{p2,d1}=1.17 J_{ss}$ was determined 
from recent experiments\cite{reuter04, reuter05}
and $\delta_{p1,p2}=0$  is assumed in 2D-EMA model.
This places in Fig. \ref{fig:phase}(a), 
for both electrons and holes, phases
$^3\Sigma$, $^2\Pi$, and $^1\Sigma$ as ground states for $N$=4, 5, 6,
respectively, where the lens represents a lens-shaped self-assembled QD, and
label ``e'', ``h'' inside the lens are symbols for electron and hole,  
respectively. As a comparison, we also show the electron ground-state phases 
for an electrostatic dot ($\sim$ 50 nm) represented by a circle.  

\subsubsection{Generic phase diagrams for atomistic approach}

We now use atomistic Coulomb and exchange energies $J$ and $K$, 
listed in Table~\ref{tab:JK-e} and~\ref{tab:JK-h} to 
recalculate the phase diagram $\{ N; \delta_{p_1,p_2}, \delta_{p_2,d_1} \}$
for electrons and holes in Figs.~\ref{fig:phase}(b) and
Figs.~\ref{fig:phase}(c), respectively. We see that the 
phase boundaries change dramatically 
for both electrons and holes.\cite{he05d} 
For example, for $N$=6 electrons, phase $^1\Sigma^*$
disappears [Fig.\ref{fig:phase}(b)],
while phase $^5\Sigma$ disappear for $N$=5 holes.

\begin{table}
\caption{Orbital occupations for $N$ holes in InAs/GaAs QDs 
of different height/base obtained from single-particle
atomistic pseudopotential 
and many-particle CI calculations. The orbital occupations
are given by the leading configurations in CI results. }
\label{tab:spin-h}
\vskip 0.2cm
\begin{tabular}{p{1.8 cm}p{.8cm} p{.8cm} p{.8cm} p{.8cm} p{1.cm} p{1.cm}}
%{l|llllll}
\hline\hline
Height/Base (nm)  & $N$=1  & $N$=2 & $N$=3 & $N$=4 
& $N$=5  & $N$=6 \\
\hline
2.5/20  & $s^1$  & $s^2$  &  $s^2p_1^1$  &  $s^2p_1^2$  &  $s^2p_1^2d_1^1$  &
  $s^2p_1^2d_1^2$\\
3.5/20  &  $s^1$  & $s^2$  &  $s^2p_1^1$  &  $s^2p_1^2$ &  $s^2p_1^2p_2^1$  &
  $s^2p_1^2p_2^2$\\
2.5/25  &  $s^1$  & $s^2$  &  $s^2p_1^1$  &  $s^2p_1^2$  &  $s^2p_1^2d_1^1$  &
   $s^2p_1^2d_1^2$\\
3.5/25  & $s^1$  & $s^2$  &  $s^2p_1^1$  &  $s^2p_1^2$  &  $s^2p_1^2d_1^1$  &
  $s^2p_1^2d_1^2$\\
2.5/27.5  &  $s^1$  & $s^2$  &   $s^2p_1^1$  &  $s^2p_1^2$  &  $s^2p_1^2p_2^1$  &
   $s^2p_1^2p_2^1d_1^1$\\
3.5/27.5  & $ s^1$ &  $s^2$  &  $s^2p_1^1$  &  $s^2p_1^2$  &   $s^2p_1^2d_1^1$  &
  $s^2p_1^2d_1^2$\\
\hline\hline
\end{tabular}
\end{table}

\subsubsection{Ground states for specific dots in the atomistic approach}

We next use $\delta_{p_1,p_2}$, $\delta_{p_2,d_1}$ and $J_{ss}$ from 
Table~\ref{tab:single} and \ref{tab:JK-e} to decide the ground states
for electrons in self-assemble InAs/GaAs QDs. For example, 
for a InAs/GaAs dot with 2.5 nm in height and 20 nm in base,
we have 
$\delta_{p_1,p_2}/J_{ss} \sim$ 0.1 and $\delta_{p_2,d_1}/J_{ss} \sim$ 3.
These parameters give the ground-state phase $^2\Pi$ for 5 electrons, 
and the ground-state phase $^1\Sigma$ for 6 electrons.
As we see, even though the phase boundaries calculated from atomistic theories 
are very different from those calculated from 2D-EMA model, 
the ground-state configurations are remain the same as those predicted 
by 2D-EMA model.
This is because in the phase diagrams,
the coordinates of the electrons in self-assembled dots
are far away from other competing phases.
For $N$=4 electrons, the high spin state phase $^3\Sigma$
(Hund's rule) and low spin state $^1\Delta$ are relatively close
in the phase diagram.
The intrinsic 
$p$-level splittings in the lens-shaped dots are about 1 - 4 meV 
(Table~\ref{tab:single}), which are smaller than the
electron-electron exchange energies ($\sim$ 5 mV), 
and therefore the ground state is the high-spin phase $^3\Sigma$. However, if 
the dot shape is elongated, adding additional $p$-level splittings, 
the ground state could be low-spin phase $^1\Delta$. Actually there are
some experimental evidences showing that the ground state of 4 electrons
could be a low spin state.\cite{miller97,wibbelhoff05}

In contrast to electrons, holes have $\Delta \epsilon_h \le J_{ss}$, and
large $p_1$-$p_2$ splitting, small $p_2$-$d_1$ splitting
(Fig.~\ref{fig:eng-level}),
which place the 
holes in a different region in the phase diagrams than electrons, where
there are more competing phases.  
For example, for the dot with base $b$= 20 nm, and height $h$=2.5 nm,
we have $\delta_{p_1,p_2}/J_{ss} \sim$ 0.4 
and $\delta_{p_2,d_1}/J_{ss} \sim$ 0.17. These relationships 
give, for this dot,
the ground states $^3\Sigma$, $^2\Delta$
and $^1\Sigma^*$ for $N$=4, 5, and 6 holes, respectively,
showing a nontrivial 
hole charging pattern that breaks the Aufbau principle.\cite{he05d}
We also list the ground state configurations 
for dots of different geometries
in Table~\ref{tab:spin-h}.
As we see, in most of the cases, the ground states are
still $^3\Sigma$, $^2\Delta$
and $^1\Sigma^*$ as shown in Fig.~\ref{fig:phase}(c).
However, there are some exceptions for very tall dots or very flat dots.
For very flat dots (e.g., the $h$/$b$ =2.5/27.5 nm dot), 
the $p_1$-$p_2$ energy splitting (5.8 meV)
is much smaller than the $p_2$-$d_1$ energy splitting (9.4 meV),
These parameters place the ground state of $N$= 5 holes in phase $^2\Pi$ 
and that of 6 holes in phase $^3\Pi$ of Fig.~\ref{fig:phase}(c).
For very tall dots (e.g., the $h$/$b$ = 3.5/20 nm dot), 
the interfacial hole localization changes the hole-hole
Coulomb integrals dramatically, and thus change the phase boundaries
in  [Fig.~\ref{fig:phase}(c)].
We find the ground states of the dot with $h$/$b$ = 3.5/20 nm
are $^2\Pi$ and $^1\Sigma$ for 5 holes and 6 holes, respectively.

\subsubsection{Effects of configuration interaction}

If we use CI instead of single-configuration approximation,
the ground states are superpositions of different configurations,
but the leading CI configurations have a significant weight, 
being 79\%, 71\% and 64\% for 4, 5, and 6 holes, respectively, in the
$h$/$b$ = 2.5/20 nm dot. 
Since the weights of leading configurations are significantly
larger than other configurations,
we are justified in using leading configurations
to represent graphically the ground states.
It is worth noting that for very large electrostatically
confined dots, 
where single-particle energy spacing
$\delta \epsilon/J_{ss} \ll 1$ 
(down-left corner of the phase diagrams of 
Fig.~\ref{fig:phase}), the ground
state mixes large number of configurations, which have no 
significant leading configurations,
and are therefore in strongly correlated states
\cite{yannouleas03}  that are not discussed here.
 
\begin{table*}
%\newpage
\caption{Summary of atomistically calculated electron addition energies
$\Delta_e$ and hole addition energies $\Delta_h$ (in
meV) of In$_{1-x}$Ga$_x$As/GaAs quantum dots for various dot heights,
base sizes and Ga compositions.
The experimental addition energies for electrons are extracted
from Ref.~\onlinecite{miller97}, and the addition energies 
for holes are taken from Ref.~\onlinecite{reuter05}. }
\label{tab:addition}
\vskip 0.2cm
\begin{tabular}{p{2.4cm} p{1.4cm} p{1.4cm} p{1.4cm} p{1.4cm} p{1.4cm} 
p{1.4cm}p{1.4cm}p{1.4cm}p{1.4cm}p{1.4cm}}
%{ldddddddddd}
\hline\hline
Height (nm)  & 2.5 & 3.5 & 2.5&3.5&3.5&3.5&3.5 & 2.5 &
35 &\\
Base (nm)   & 20  & 20 & 25 &25  &25 &25 &25  & 27.5 & 27.5 \\
Ga comp.  &  0     &   0    &  0     &0 & 0.15  & 0.3    & 0.5  & 0  & 0 
&Exptl.\\
\hline
$\Delta_e(1,2)$ &  26.3 & 24.6 & 23.3 & 21.8 & 20.5 & 18.5 & 15.2 & 22.1 & 20.5
& 21.5  \\
$\Delta_e(2,3)$ &  89.5 & 84.4  & 71.8 & 67.6 & 61.2 & 53.2 & 40.3 & 65.2 &
60.8 &$\sim$ 57\\
$\Delta_e(3,4)$ & 19.7  & 20.3  & 16.7 & 16.5 & 15.9 & 12.7 & 10.6 & 15.4 &
15.5 & 11.4 \\
$\Delta_e(4,5)$ & 21.5  & 19.5  & 20.1 & 18.5 & 16.6 & 16.5 & 12.9 & 19.6 &
17.5 & 21.0 \\
$\Delta_e(5,6)$ & 19.7  & 20.3  & 16.7 & 16.5 & 15.8 & 12.4 & 10.3 & 15.4 &
15.4 &12.2 \\
\hline
$\Delta_h(1,2)$ & 24.1 & 19.0 & 21.9 & 17.5 & 18.3 & 18.4 &17.8 & 21.0 & 16.7&
23.9\\
$\Delta_h(2,3)$ & 28.7& 21.7  & 27.2 & 21.2 & 23.0 & 23.6 & 22.6& 26.4 & 20.6
& 34.2 \\
$\Delta_h(3,4)$ & 18.1& 16.9  & 16.4 & 15.2 & 15.4& 15.4 & 15.0& 15.6 & 14.5& 17.1\\
$\Delta_h(4,5)$ & 26.4& 21.6  & 25.4 & 20.8 & 23.2& 21.7 & 19.3 & 23.8 & 20.5&
23.2\\
$\Delta_h(5,6)$ & 17.1& 16.1  & 15.3 & 14.4 &14.5& 16.0 & 15.7& 15.5 & 13.7 & 15.0\\
\hline\hline
\end{tabular}
\end{table*} 

\subsection{Calculated charging and addition energies}
\label{sec:addE}

Once we determined the ground state configurations, 
we can calculate the total energies using Eq.~(\ref{eq:tot-eng}).
We calculate the ground state total energies for up to
6 electrons/holes for each dot. 
For electrons, in the CI approach
we used 6 single-particle electron levels ($s$, $p_1$, $p_2$, $d_1$,
$d_2$, $d_3$) to construct all possible Slater determinants, 
while for holes, we used 8  single-particle hole levels. 
The total number of Slater determinants for 6 electrons is 924.
For 6 holes, the total number of determinants is 8008. 
We plot in Fig. \ref{fig:CI-converge} the CI total 
energies for the ground state of 6 carriers {\it vs.} 
number of single-particle states 
included in the CI expansions. The total energies converge to about 1 meV if 6
single-particle states are used for electrons and 8 states are
used for holes.

\begin{figure}
\includegraphics[width=3.2in,angle=0]{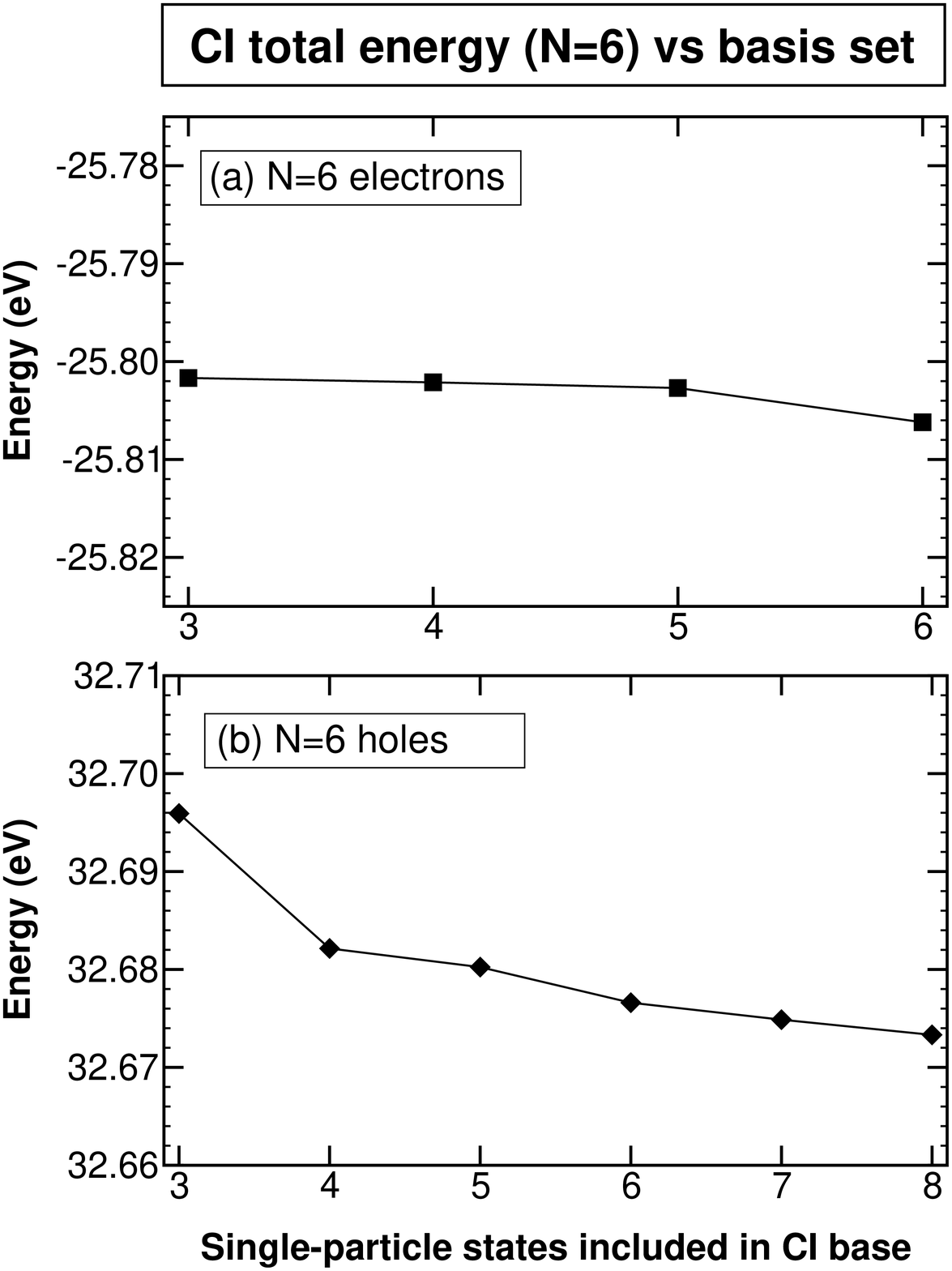}
\caption{ CI total-energy convergency test for (a) six-electron and (b) 
six-hole states with increasing number of single-particle orbitals 
in the bases. 
The test is done for a pure InAs/GaAs dot, with 
base $b$=25 nm, and height $h$=3.5 nm. 
}
\label{fig:CI-converge}
\end{figure}

\begin{figure}
\includegraphics[width=2.6in,angle=0]{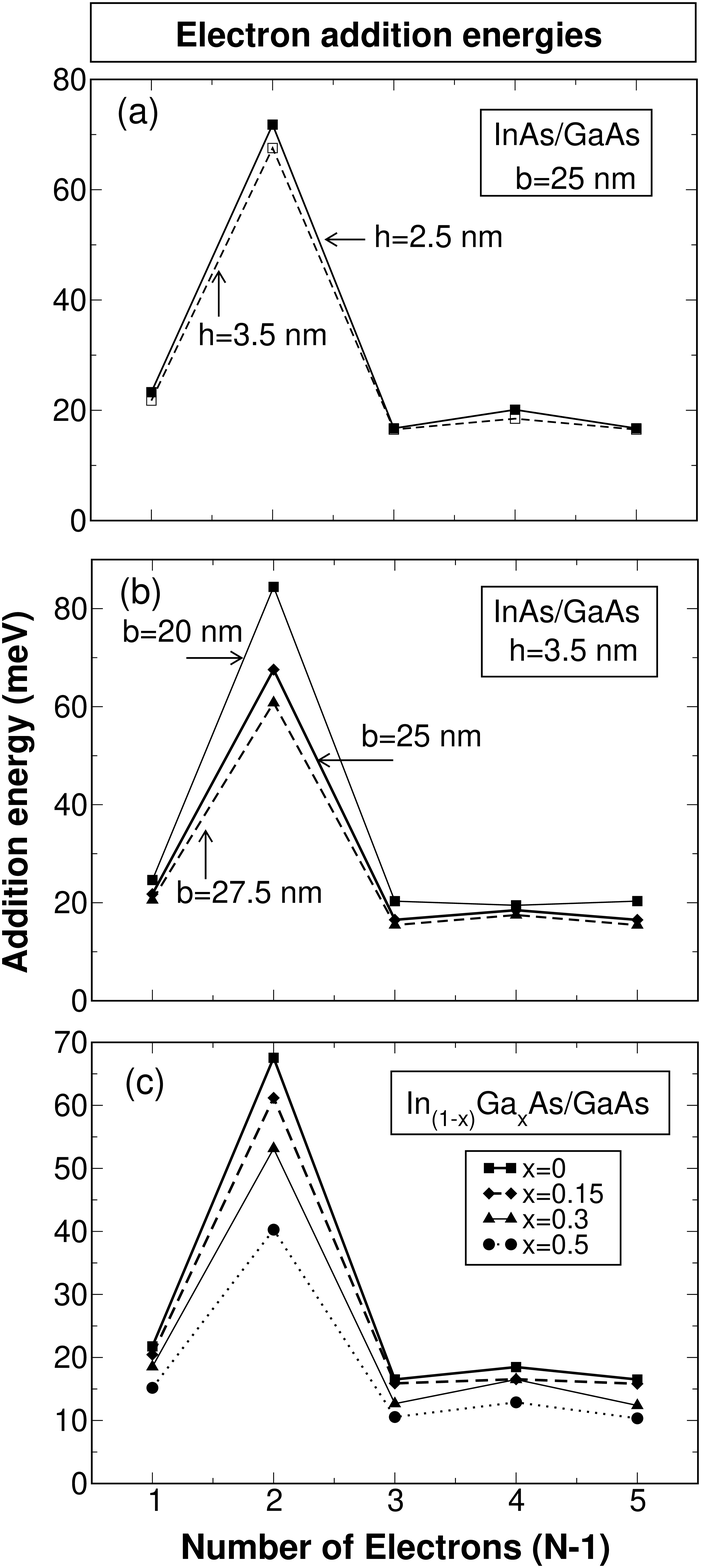}
\caption{ Electron addition energies of 
(a) pure InAs/GaAs dots with base $b$=25 nm and 
height $h$=2.5 nm and 3.5 nm; (b) pure InAs/GaAs dots
with height $h$=3.5 nm and base $b$=20 nm, 25 nm 
and 27.5 nm; and (c) In$_{1-x}$Ga$_{x}$As/GaAs dots 
with height $h$=3.5 nm and base $b$=25 nm
at $x$=0, 0.15, 0.3 and 0.5.
}
\label{fig:add-e}
\end{figure}

\begin{figure}
\includegraphics[width=2.6in,angle=0]{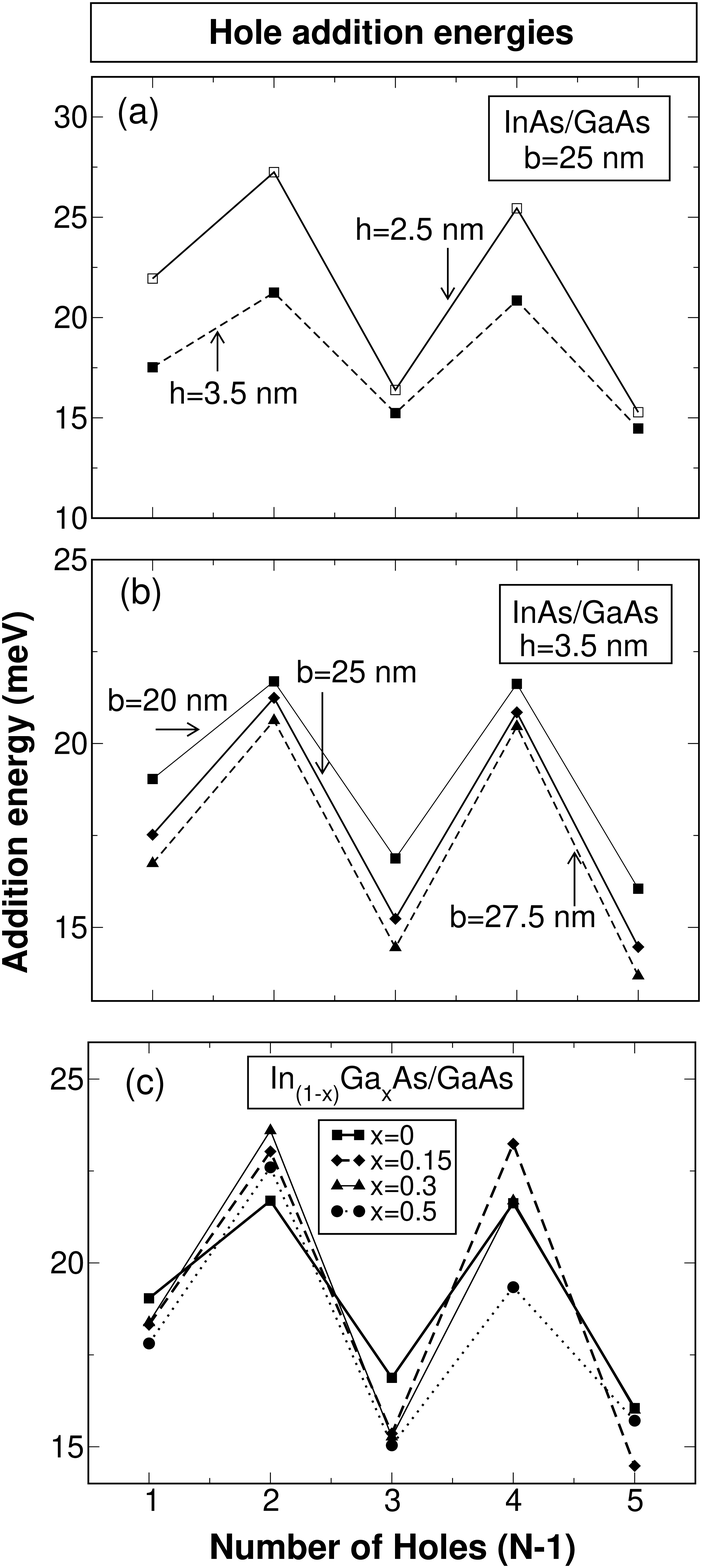}
\caption{ Hole addition energies of 
(a) pure InAs/GaAs dots with base $b$=25 nm and 
height $h$=2.5 nm and 3.5 nm; (b) pure InAs/GaAs dots
with height $h$=3.5 nm and base $b$=20 nm, 25 nm and 27.5 nm;
and (c) In$_{1-x}$Ga$_{x}$As/GaAs dots 
with height $h$=3.5 nm and base $b$=25 nm
at $x$=0, 0.15, 0.3 and 0.5.
}
\label{fig:add-h}
\end{figure}

The charging energies and addition energies are calculated using 
Eq.~(\ref{eq:charging}) and Eq.~(\ref{eq:addition}). 
The addition energies calculated from CI calculations
give results that are about 1 meV different from 
single-configuration results for electrons,
and 1 - 3 meV different for holes. 
The addition energies
are summarized in Table~\ref{tab:addition}.
Experimentally, the charging energies of electrons~\cite{drexler94,
fricke96,miller97} and holes \cite{reuter04, reuter05} in
self-assembled QDs 
are usually measured via capacitance spectroscopy, using 
gated structures.\cite{footnote2}
The (In,Ga)As/GaAs dots used in the electron charging experiments were roughly
estimated to be 
7 nm in height and 20 nm in base in Refs.~\onlinecite{fricke96,miller97}.
The measured $\Delta_e(1,2) \sim$ 23 meV in Ref.~\onlinecite{fricke96}
and $\sim$ 21.5 meV in Ref.~\onlinecite{miller97}, 
which, as shown in Sec.~\ref{sec:background}, 
equals roughly $J_{ss}^{(ee)}$, the electron-electron Coulomb 
interaction of $s$ orbitals. 
The charging energies between $s$ and $p$ levels,
$\Delta_e(2,3)$ was estimated to be 57.0 meV in 
Ref.~\onlinecite{miller97}.
The average addition energies between $p$-states 
are estimated to be 18 meV in Ref.~\onlinecite{fricke96},
and 14 meV in Ref.~\onlinecite{miller97}.
Furthermore, in Ref.~\onlinecite{miller97}, 
$\Delta_e(4,5)$ is almost twice as large as 
$\Delta_e(3,4)$ and  $\Delta_e(5,6)$, which might be related to 
the breakdown of Hund's
rule as a consequence of irregular shape of the dots.
These results are listed in Table~\ref{tab:addition} and compared with 
our theoretical results.
We see that the electron addition energies of 
an (In,Ga)As/GaAs dot, with $b$/$h$ = 3.5/25 nm and Ga composition $x$=
0.15, agree very well with the above experimental results, which
show $\Delta_e(1,2)$= 20.5 meV, $\Delta_e(2,3)$= 61.2 meV, and
the average addition energies between $p$-states of about 16 meV.
  
The experimental hole addition energies are taken 
from Ref.~\onlinecite{reuter05}, which gives $\Delta_h(1,2)$=23.9 meV, 
comparable to that of
$\Delta_e(1,2)$. However, the addition energy between $s$ and $p$ orbitals,
$\Delta_h(2,3)$=34.2 meV, is significantly smaller
than $\Delta_e(2,3) \sim$ 57 meV. This result reflects that the $s$-$p$
energy spacing of holes is much smaller than that of electrons.
As seen from Table~\ref{tab:addition}, our calculated addition energies
of pure and flat (height=2.5 nm) InAs/GaAs dots agree very well with this experiment.

To study trends of addition energies for the electrons and the holes, 
we depict the electron addition energies for different dot heights
[Fig.~\ref{fig:add-e}(a)],  bases [Fig.~\ref{fig:add-e}(b)] 
and alloy compositions [Fig.~\ref{fig:add-e}(c)].
Similarly, the hole addition energies are plotted in 
Figs.~\ref{fig:add-h} (a),~\ref{fig:add-h}(b), and~\ref{fig:add-h}(c)  
for dot heights, bases and alloy compositions.
We see the following:

(i) Up to six carriers, the
{\it electron} addition energies $\Delta(N-1,N)$ 
have a single peak 
at ($N-1$)=2. The peak is due to the single-particle 
energy
gap between the $s$-shell and $p$-shell. On the other hand,  
all {\it hole} addition energies have {\it two} peaks at ($N-1$)=2 and 
($N-1$)=4 respectively,
where the first peak come from the single-particle energy gap between the 
$s$ orbital and $p_1$
orbital and the second peak is associated with the energy 
difference between the $p_1$ and higher energy
orbitals.

(ii) Electron addition energies decrease with increasing height and base
of the dots, and the hole addition energies share the same trend. However, 
the electron addition energies are more sensitive to the
base of the dots and relatively insensitive to the height. In contrast,
the hole addition energies are very sensitive to the heights of the dot,
and  relatively insensitive to the base. 
This dependence suggests that the electron wave functions 
are more 2-dimensional 
like in the QDs, 
while hole wave functions are more extended in [001] direction.
 
(iii) Electron addition energies show a simple trend for alloyed 
In$_{1-x}$Ga$_x$As/GaAs QDs, which are
consistently smaller for Ga rich dots. However, the 
hole addition energies of alloy dots show more complicated
behaviors.  The reasons of this complication are the following:
(1) alloy dots have different trends in hole single-particle 
energy level spacings (see Table \ref{tab:single});
(2) the ground state configurations might be different
for different alloy compositions.

\section{summary}
\label{sec:summary}

We systematically studied the electron/hole addition energy spectra using
single-particle pseudopotential plus many-particle CI methods. 
Considering the single-particle step, we find for {\it electrons} that
there is a shell structure and that the $p$-$p$ and $d$-$d$ splittings
are about 1 - 4 meV depending on the dot geometry.
For {\it holes}, the single-particle step reveals
large (5 - 11 meV) 
$p_1$-$p_2$ splitting and absence of a
well defined shell structure.
Considering the $e$-$e$ and $h$-$h$                        
{\it Coulomb integrals}, we find that atomistically calculated
ratios between various Coulomb integrals $J_{i,j}$ 
differ by about 20\% from those in the 2D-EMA model calculation. 
These differences lead to many-particle phase diagrams 
that differ significantly from those predicted
by the 2D-EMA model. 
In particular, the ``unusual'' hole
single-particle spectrum and Coulomb integrals
lead to  many-particle ground states
that defy the Hund's rule and the Aufubau principle for holes.
The predicted ground-state configurations and
the {\it addition energies} calculated in this
pseudopotential plus CI scheme  
compare well with experiments.

\acknowledgments
We thank G. Bester and G. A. Narvaez for fruitful discussions.
This work was funded by the U.S. Department of Energy, Office of Science,
Basic Energy Science, Materials Sciences and Engineering, LAB-17 initiative,
under Contract No. DE-AC36-99GO10337 to NREL.


\begin{thebibliography}{48}
\bibitem[{foo({\natexlab{a}})}]{footnote0}
\bibinfo{note}{For alloyed dots, the same dependence with height of the
  $s$-$-p$ spacing is find. See Ref.~\onlinecite{narvaez05}.}
\bibitem[{foo({\natexlab{b}})}]{footnote1}
\bibinfo{note}{The piezoelectric effect also contributes to the $p$-state
  splitting of the electron states, as has been shown in
  Ref.\onlinecite{bester05a}. For the splitting of the hole states, the effect
  is more subtle. We find for a pure InAs lens shaped dot with 25 nm base and
  3.5 nm height, using piezoelectric coefficients from bulk materials, that the
  hole $p$-$p$ splitting without piezoelectric effect is 9.5 meV and with the
  piezoelectric effect is 9.7 meV, using piezoelectric coefficients from bulk
  materials. The change (0.2 meV) due to piezoelectric effects is rather small.
  We therefore ignored piezoelectric effects in our calculations.}
\bibitem[{foo({\natexlab{c}})}]{footnote2}
\bibinfo{note}{The electric fields applied to the dots are typically of the
  order of 50 kV/cm for electron charging devices (as gleaned from Ref.
  \onlinecite{drexler94,fricke96,miller97}) and 100 kV/cm for hole charging
  devices.~\cite{reuter04,reuter05} We performed additional calculations to
  estimate the effect of this field on the addition energies and obtained
  changes for addition energies to be about 1 meV at 100 kV/cm . We therefore
  ignored this correction in our calculations.}
\end{thebibliography}
\end {document}